\documentclass[%
 reprint,
 amsmath,amssymb,
 aps,
]{revtex4-2}

\usepackage{graphicx}
\usepackage{dcolumn}
\usepackage{bm}

\usepackage{xcolor}
\usepackage{soul}

\begin{document}

\preprint{APS/123-QED}

\title{On the microscopic origin of Soret coefficient minima in liquid mixtures}

\author{Oliver R. Gittus}
 \email{o.gittus18@imperial.ac.uk}
\author{Fernando Bresme}%
 \email{f.bresme@imperial.ac.uk}
\affiliation{Department of Chemistry, Molecular Sciences Research Hub, Imperial College London,\\ London W12 0BZ, United Kingdom.}%

\date{23 July 2022}

\begin{abstract}
Temperature gradients induce mass separation in mixtures in a process called thermodiffusion and quantified by the Soret coefficient. The existence of minima in the Soret coefficient of aqueous solutions was controversial until fairly recently, where a combination of experiments and simulations provided evidence for the existence of this physical phenomenon. However, the physical origin of the minima and more importantly its generality, e.g. in non-aqueous liquid mixtures, is still an outstanding question. 
We report the existence of a minimum in liquid mixtures of non-polar liquids modelled as Lennard-Jones mixtures, demonstrating the generality of this phenomenon. 
The Soret coefficient minimum originates from a coincident minimum in the thermodynamic factor, and hence denotes a maximimzation of non-ideality mixing conditions.
We explain the microscopic origin of this effect in terms of the atomic coordination structure of the mixtures.
    
\end{abstract}

\maketitle


{\it Introduction:}
Thermal gradients induce the transport of
colloids in suspension (thermophoresis) and concentration gradients in liquid mixtures and solutions (thermodiffusion)~\cite{duhr2006,Duhr19678,rasuli2008,wurger2008,Wiegand2004_SoretReview,piazza2008,Kohler2016_ThermalDiffusionReview}. Thermodiffusion was discovered in the 19$^{th}$ century: first observed by Ludwig~\cite{ludwig1856}, and later systematically investigated by Soret\cite{soret1879}. The Soret coefficient, $S_T$, measures the mass separation of mixtures and suspensions in a thermal field. It is becoming a central property to characterise the non-equilibrium response of soft matter to thermal fields, for example, to quantify thermal transport in micro-scale devices that analyse the interactions of proteins and small molecules in biological liquids.~\cite{wienken2010}

Experimental and computational studies have advanced significantly in recent years, but several outstanding questions remain. One such question is the microscopic origin of the forces driving the phenomenology observed in thermodiffusion measurements. Aqueous solutions feature a particularly rich phenomenology. For example, experiments and simulations report changes in the thermophilicity of alkali halide salt solutions with temperature. The solutions are thermophobic ($S_T>0$) at high temperatures and thermophilic ($S_T<0$) at low temperatures\cite{alexander1954,Gaeta1982_SoretSaltSolutions,Romer2013_SoretSaltSolutions}.  
Interestingly, Gaeta et al.~\cite{Gaeta1982_SoretSaltSolutions} reported minima in the Soret coefficients of {NaCl$_\mathrm{(aq)}$} and {KCl$_\mathrm{(aq)}$} as a function of concentration. These experiments were performed with thermogravitational columns, and the minima could not be reproduced using state-of-the-art thermal diffusion forced Rayleigh scattering techniques, which circumvent convection effects\cite{Romer2013_SoretSaltSolutions}. Hence the $S_T$ minimum has remained controversial for many years. However, this situation has changed with recent experiments and computer simulations of {LiCl$_\mathrm{(aq)}$}, which support the existence of a minimum in the Soret coefficient~\cite{Colombani1999_SoretLiCl,DiLecce2017_SoretLiClHeatOfTransport}. 
Very recently, minima at high concentrations ($\sim 2$~M) were observed for thiocyanate (NaSCN$_\mathrm{(aq)}$ and KSCN$_\mathrm{(aq)}$) and acetate (CH$_3$COOK$_\mathrm{(aq)}$) salt solutions~\cite{Mohanakumar2021_ThermodiffusionAqueousSolutions,Mohanakumar2022}, giving further impetus to the investigation of the physical origin of the Soret coefficient minima. 
We note that understanding the physical variables controlling the minimum is potentially relevant to explain the thermoelectric response of aqueous solutions as well (see ref.\cite{DiLecce2017_SoretLiClHeatOfTransport}).

In addition to aqueous electrolyte solutions, minima in the Soret coefficient with composition were observed in mixtures of polar fluids:  ethanol/water\cite{Wiegand2004_SoretReview,Koniger2009_SoretMeasurements,Zhang2006_EthanolWater}, dimethyl-sulfoxide/water~\cite{Ning2006_SoretPolarSolvents} and acetone/water~\cite{Ning2006_SoretPolarSolvents,Cabrera2009_SoretAcetoneWater}. In all these systems (electrolyte solutions and polar fluid mixtures), one of the components is water. 
This observation might suggest that the Soret coefficient minima are interlinked with water as a solvent and, therefore, its specific thermal transport properties.
Indeed, molecular simulations of atomistic (non-polar) Lennard-Jones (LJ) binary mixtures at supercritical conditions do not offer evidence for the existence of minima in the Soret coefficient with composition~\cite{Artola2007_SoretChemContr}. However, some experiments of non-polar or weakly polar liquid mixtures reported maxima/minima in the Soret coefficient (e.g. cyclohexane/cis-decaline) and in some cases, accounting for an extrapolation to infinite dilution, a weak extrema can be inferred 
(e.g. {toluene/1,3-dichlorobenzene})~\cite{Hartmann2014_ThermophobicitySoret}. 
That work made no attempt to explain the microscopic origin of the extrema in $S_T$, but crucially highlights the importance of the thermodynamic factor, $\Gamma$, a key quantity determining the heat of transport~\cite{Agar1989,DiLecce2017_SoretLiClHeatOfTransport}.

To investigate the existence of Soret coefficient minima in non-polar mixtures, and to probe the microscopic origin of such minima, we have performed computer simulations of the simplest liquid binary mixture, modelled with the Lennard-Jones model, which accounts for dispersion interactions. 
Advancing the discussion below, we show for the first time that a minimum in the Soret coefficient at a specific composition, and constant temperature and pressure, can be observed in simple non-polar liquid mixtures, hence showing that the minimum in the Soret coefficient is a completely general physical phenomenon. We
demonstrate that the minimum in the systems investigated here emerges from a balance of the mutual diffusion coefficient and the thermal diffusion coefficient, which vary in the opposite direction upon changing the composition of the mixtures. 
Furthermore, we correlate the minimum to the thermodynamic factor and changes in the solvation structure of the particles. 
Finally, we show that existing theories of thermodiffusion either do not reproduce the minimum and/or predict inaccurate $S_T$ values.

{\it Methods: } We investigate the non-equilibrium behaviour of liquid-liquid binary Lennard-Jones mixtures at constant temperature $T$, pressure $P$, and different mole fractions $x_1 = N_1/ (N_1+N_2)$, where $N_i$ is the number of particles of type $i$.
Inter-particle interactions were modelled using the LJTS potential, which is the LJ potential
$\mathcal{V}_{ij}^\mathrm{LJ}(r) = 4\epsilon_{ij} [ (\sigma_{ij}/r)^{12} - (\sigma_{ij}/r)^{6}]$
truncated and shifted at a cutoff radius of $r_c = 2.5\sigma$,
$\mathcal{V}_{ij}^\mathrm{LJTS}(r) = (\mathcal{V}_{ij}^\mathrm{LJ}(r) - \mathcal{V}_{ij}^\mathrm{LJ}(r_c)) \theta(r_c-r) $
with $\theta$ being the Heaviside step function. 
All the particles have the same diameter and mass, but the interactions between of particles of type ``1'' or ``2'' are different.
The parameters $\epsilon = \epsilon_{22}$ and $\sigma = \sigma_{11} = \sigma_{22}$ together with the mass of each particle $m_1 = m_2 = m$ define the usual LJ units. The energy scale was defined in terms of the high-boiling component, with $\epsilon_{11} / \epsilon_{22} = 0.6$. The Lorentz-Berthelot combining rules were used for the inter-species interaction parameters: $\epsilon_{12} = \sqrt{\epsilon_{11} \epsilon_{22}} = \sqrt{0.6} \epsilon$ and $\sigma_{12} = (\sigma_{11} + \sigma_{22}) / 2 = \sigma$. 
Experiments of liquid-liquid mixtures have reported extrema
in $S_T$, therefore we target subcritical conditions for the LJ mixtures. Liquid-liquid mixtures were modelled at temperature {$T = 0.62$~{$\epsilon k_\mathrm{B}^{-1}$}} ($k_\mathrm{B}$ is the Boltzmann constant) and pressure {$P = 0.46$~{$\epsilon \sigma^{-3}$}}, below the critical point\cite{Thol2015_LJTSEOS} for both species. The entire $0 \leq x_1 \leq 1$ composition domain is therefore expected to be subcritical. We note that along this isobar-isotherm, the thermodynamically stable phase for $x_1 = 0$ is a solid and we estimate the liquidus to be at $x_1 \approx 0.2$ (see sec.~1.1 in the SI for further discussion). 
Simulations below this mole fraction correspond to a metastable liquid-liquid mixture. Nevertheless, as we will show below, the observed $S_T$ minimum is safely within the liquid-liquid mixture portion of the phase diagram.

We performed a variety of equilibrium molecular simulations (EMS), molecular dynamics (MD) and Monte Carlo methods, as well as non-equilibrium molecular dynamics (NEMD) simulations to calculate $S_T$ and related quantities. From NEMD, $S_T$ was evaluated at the stationary state characterised by zero net mass flux ($\bm{J}_1 = 0$) as
\begin{equation}\label{eq:STstationary}
    S_T = - \left( \frac{1}{w_1 w_2} \frac{\nabla w_1}{\nabla T} \right)_{\bm{J}_1 = 0}
    = - \left( \frac{1}{x_1 x_2} \frac{\nabla x_1}{\nabla T} \right)_{\bm{J}_1 = 0}
\end{equation}
where $x_i$ and $w_i$ are the mole and mass fractions of species $i$ (related by $w_i = x_i m_i / (x_1 m_1 + x_2 m_2)$ where $m_i$ is the mass of species $i=1,2$). EMS methods calculate,
\begin{equation}\label{eq:STdef}
S_T \equiv \frac{D_T}{D_{12}}
\end{equation}
\noindent
from~\cite{GrootMazurNET} 
\begin{equation}
    D_{12} = \frac{ L_{11}}{ \rho (1-w_1) T } \left(\frac{\partial \mu_{s,1}}{\partial w_1} \right)_{P,T} 
\end{equation}
\begin{equation}
    D_T = \frac{ L_{1q}^{\prime} }{ \rho w_1 (1-w_1) T^2}
\end{equation}
where $D_{12}$ is the mutual diffusion coefficient and $D_T$ is the thermal diffusion coefficient.
Onsager's phenomenological coefficients $L_{\alpha \beta}$ were calculated from MD simulations in the \textit{NVE} ensemble, using the Green-Kubo (GK) integral formulas and taking into account the enthalpy terms for the primed coefficient $L_{1q}^{\prime}$ (see the SI). The chemical potential $\mu_1$, and subsequently the specific chemical potential $\mu_{s,1} = \mu_1/m_1$, was calculated from MD simulations using a free energy perturbation (FEP) method at constant $T$ and $P$. $(\partial \mu_1 / \partial x_1)_{P,T}$ was then calculated from the numerical derivative of $\mu_1$. $(\partial \mu_1 / \partial x_1)_{P,T}$ was also obtained from Kirkwood–Buff solution theory via two different methods of evaluating the Kirkwood-Buff integrals (KBIs): (1) from particle number fluctuations in grand canonical Monte Carlo (GCMC) simulations and 
(2) from the extrapolation to infinite system size of finite-volume KBIs\cite{Kruger2013_FiniteVolumeKBI}, which were in turn calculated  using radial distribution functions (RDFs) from MD simulations in the \textit{NVT} ensemble.
Thus, three EMS methods were used to calculate $(\partial \mu_1 / \partial x_1)_{P,T}$ and subsequently the thermodynamic factor: FEP, KBI(RDF) and KBI(GCMC). Combining these with the GK calculations for the phenomenological coefficients give three corresponding ``equilibrium'' routes to $S_T$, GK+FEP, GK+KBI(RDF) and GK+KBI(GCMC), in addition to NEMD simulations.    
In addition, other thermophysical properties such as the equation of state, diffusion coefficients and viscosities were calculated from equilibrium-MD simulations in the \textit{NVE}, \textit{NVT} and \textit{NPT} ensembles. Details about these simulations and calculations are given in the SI. All simulations were performed using the software package LAMMPS~\cite{LAMMPS} (v. 3 March 2020).

\begin{figure*}[!ht]
    \centering
    \includegraphics[width=0.8\textwidth]{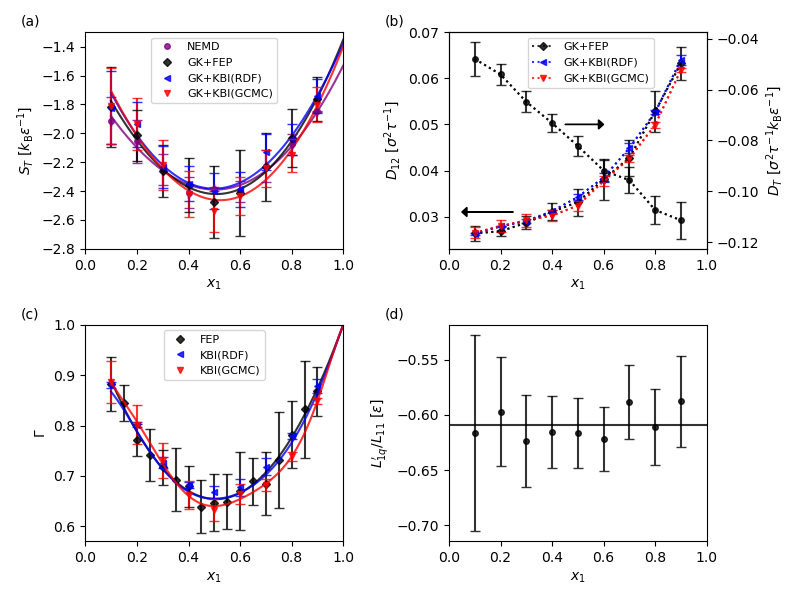}
    \caption{The Soret coefficient and related properties of the LJ mixture as a function mole fraction $x_1$. (a) The Soret coefficient $S_T$; the solid lines show cubic functions fit to the $S_T(x_1)$ data. (b) The mutual diffusion coefficient $D_{12}$ (left axis) and thermal diffusion coefficient $D_T$ (right axis). (c) The thermodynamic factor $\Gamma$; the solid lines show polynomial functions fit to $\Gamma(x_1)$ with the infinite-dilution constraint $\Gamma(1)=1$.  (d) The ratio of phenomenological coefficients $L_{1q}^{\prime}/L_{11}$; the solid line shows the weighted arithmetic mean. 
    }
    \label{fig:main}
\end{figure*}

\textit{ Results \& discussion:}
{Fig.~\ref{fig:main}(a)} contains the main result of this letter: $S_T$ features a minimum as a function of composition at 
{$x_1^{\mathrm{min}(S_T)} \sim 0.5$}. Fitting cubic functions to the data give {$x_1^{\mathrm{min}(S_T)} = 0.5 \pm 0.1$} for all four methods: NEMD, GK+FEP, GK+KBI(RDF) and GK+KBI(GCMC).
The thermodiffusion response at the minimum is significantly enhanced with respect to diluted mixtures, by {$\sim$30-40\%} relative to $x_1 = 0.1, 0.9$, and by {$\sim$60-80\%} when compared to the extrapolated value of $S_T$ at $x_1 = 1$. 
For all compositions, $S_T < 0$ which indicates that species 1 (the low-boiling component) is thermophilic and preferentially collects in the hot region. 
The $S_T$ values calculated from NEMD and all three EMS methods are in excellent agreement: all values agree within their statistical uncertainties.
Accepting that the Green-Kubo integrals are sufficiently well converged (as shown in sec.~2.4 of the SI) the agreement indicates that the NEMD results are within the linear regime. This is expected since 
{$\nabla T T^{-1} \sigma \lesssim 0.007 \ll 1$.} 
Further verification of linear response is provided in {sec.~2.2.2 of the SI} for two mole fractions ($x_1 = 0.5, 0.9$).

$S_T$ is determined by $D_{12}$ and $D_T$ (see {eq.~\ref{eq:STdef}}), which monotonically increase and decrease with $x_1$, respectively (fig.~\ref{fig:main}(b)).
Thus, the $S_T$ minimum arises from a balance of $D_T$ and $D_{12}$, as opposed to being carried through only by one of the transport coefficients.

The Soret coefficient $S_T$ can be written in terms of the phenomenological coefficients and the thermodynamic factor $\Gamma$, as~\cite{GrootMazurNET}
\begin{equation}\label{eq:STphem}
    S_T = \frac{L_{1q}^{\prime}}{L_{11} T w_1} \left(\frac{\partial \mu_{s,1}}{\partial w_1} \right)_{P,T}^{-1} =  \frac{1}{k_\mathrm{B} T^2} \frac{L_{1q}^{\prime}}{L_{11}} \frac{m_1}{\Gamma}
\end{equation}
\noindent 
 \begin{equation}
    \Gamma = \frac{x_1}{k_\mathrm{B} T} {\left( \frac{\partial \mu_1}{\partial x_1} \right)_{P,T}}
\end{equation}
\noindent
where $m_1 = m_2 \Leftrightarrow x_1 = w_1$.
The analysis of the different contributions to the 
RHS of {eq.~\ref{eq:STphem}} offers microscopic insight into the mechanisms determining the minimum in $S_T$. 
We find that the ratio $L_{1q}^{\prime}/L_{11}$ is essentially constant for all compositions (fig.~\ref{fig:main}(d)). 
This suggests that the minimum arises from the $w_1 (\partial \mu_1 / \partial w_1)_{P,T}$ term. 
As shown in {fig.~\ref{fig:main}(c)}, $\Gamma$ features a distinctive minimum at $x_1^{\mathrm{min}(\Gamma)} \sim 0.5$. 
{All three EMS methods (FEP, KBI(RDF) and KBI(GCMC)) predict $x_1^{\mathrm{min}(\Gamma)}$ values in good agreement with each other.}
\textit{\it The minimum in $S_T$ is connected to the minimum in the thermodynamic factor, and therefore the minimum signals the 
composition at which the mixture features the largest non-ideality, $\max |1 - \Gamma|$}. 
We note that the $S_T$ and $\Gamma$ minima coincide nearly exactly, due to the weak composition dependence of {$L_{1q}^\prime / L_{11}$}. 

\begin{figure*}[!ht]
    \centering
    \includegraphics[width=0.95\textwidth]{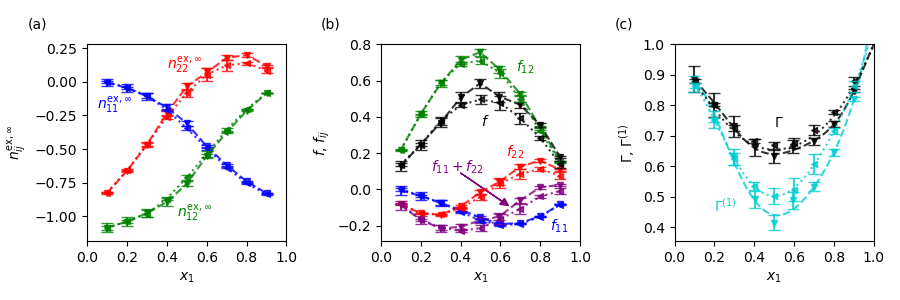}
    \caption{Analysis of the thermodynamic factor $\Gamma$ as a function of mole fraction $x_1$ using Kirkwood-Buff theory. (a) Excess coordination numbers $n^\mathrm{ex,\infty}_{ij}$ and (b) related functions $f$ and $f_{ij}$. (c) $\Gamma$ and its first-order approximation $\Gamma^{(1)}$ (see main text for additional details).
    Symbols: sideways triangles and dotted lines ($\cdot \cdot \triangleleft \cdot \cdot$) denote the KBI(RDF) data; downwards triangles (-{}-$\triangledown$-{}-) and dashed lines deonote the KBI(GCMC) data. 
    In (c), dotted and dashed lines show polynomial functions fit to the {KBI(RDF)} and {KBI(GCMC)} data respectively. Fits to $\Gamma(x_1)$ were performed with the infinite-dilution constraint $\Gamma(0)=1$. 
    }
    \label{fig:nex_analysis}
\end{figure*}

The results presented above indicate that the non-ideal contribution dominates at conditions near the minimum. This can be visualised by splitting $\Gamma$ into its ideal (id) and excess (ex) parts {(see sec.~1.3 in the SI)}, showing as expected that $\Gamma^\mathrm{ex}$, 
is responsible for the minimum in $\Gamma$.
To gain insight into the microscopic origins of the minimum in $\Gamma$ and $S_T$, we turn to Kirkwood-Buff theory, which connects $\Gamma$ to the structural properties of the binary mixture,
\begin{equation}\label{eq:TF_KBI}
    \Gamma = 1 - \frac{x_1 x_2 \rho_N (G_{11} + G_{22} - 2G_{12})}{1 + x_1 x_2 \rho_N (G_{11} + G_{22} - 2G_{12})} 
\end{equation}
where $\rho_N$ is the total number density of the mixture.
The Kirkwood-Buff integral (KBI) $G_{ij}$ is defined as the spatial integral over $g_{ij}^{\mu V T}(r) - 1$, and quantifies the excess (or deficiency) of species $j$ around $i$. $g_{ij}^{\mu V T}(r)$ is the pair correlation function in the grand canonical ensemble. 
The KBIs can be expressed in terms of the 
excess coordination numbers $n^\mathrm{ex}_{ij}(r^{\prime}) = n_{ij}(r^{\prime}) - n^\mathrm{id}_{ij}(r^{\prime})$ where $n_{ij}$ is the total coordination number, obtained from an integral over the pair correlation function, and $n^\mathrm{id}_{ij} = (4 \pi /3) \rho_{N,j} r_c^{\prime 3}$ is the ideal correlation number. $\rho_{N,j}$ is the number density of species $j$.
Hence, the KBIs are given by
$\rho_{N,j} G_{ij} = \lim_{r^{\prime}\to\infty} n^\mathrm{ex}_{ij}(r^{\prime}) = n^\mathrm{ex, \infty}_{ij}$, and the thermodynamic factor by
$\Gamma = (1+f)^{-1}$
where $f = (1-x_1) n^\mathrm{ex,\infty}_{11} + x_1 n^\mathrm{ex,\infty}_{22} - 2 x_1 n^\mathrm{ex,\infty}_{12} = f_{11} + f_{22} + f_{12}$.

In order to disentangle the contributions from $n_{11}^\mathrm{ex,\infty}$, $n_{22}^\mathrm{ex,\infty}$ and $n_{12}^\mathrm{ex,\infty}$ we take the first-order approximation to the thermodynamic factor, $\Gamma^{(1)} = 1 - f$. As shown in fig.~\ref{fig:nex_analysis}(c), $\Gamma^{(1)}$ results in underestimations of {1-35\%} across the range of compositions, with larger errors for more non-ideal mixtures, but nevertheless provides insight into the relative importance of the $f_{ij}$ terms. 
$f_{12}$ features a maximum at $x_1 \sim 0.5$ ({fig.~\ref{fig:nex_analysis}(b)}), indicating that the cross-species contribution is responsible for the minimum in $\Gamma^{(1)}$ and $\Gamma$. 
{$|(f_{11}+f_{22})/f_{12}|=0$-$0.41$} 
 making the cross-species contribution much more significant, and {$\sim$3-4} times larger in the region of the $\Gamma^{(1)}$ and $\Gamma$ minima. Consequently, the phenomenology of $\Gamma^{(1)}$ and $\Gamma$ are primarily determined by $f_{12}$. 
\textit{Thus, the composition dependence of $n_{12}^\mathrm{ex,\infty}$ (fig.~\ref{fig:nex_analysis}(a)), which represents a net depletion of species $2$ around $1$ relative to the ideal state, and increases monotonically with $x_1$, is the primary microscopic origin of the minimum in $\Gamma$ and therefore $S_T$.}

We have shown above that the Soret coefficient of a simple binary mixture of non-polar liquids features a minimum as a function of composition, and linked this minimum to non-ideal effects, which result in a minimum in the thermodynamic factor and distinctive changes in the structural properties of the mixture, as reflected in the KBIs. 
Now we examine the accuracy of existing theoretical approaches to predict the $S_T$ minimum reported above. We note that existing theoretical models do not accurately predict $S_T$ in general.~\cite{Eslamian2009_SoretModelReview,Kempers1989_SoretModel,Shukla1998_SoretModel,Kempers2001_SoretModel,Mariana2003_SoretModelEval,Artola2008_SoretModel,Hoang2022_SoretModels} In some cases, even the sign of $S_T$ is not predicted correctly.\cite{Kempers1989_SoretModel,Shukla1998_SoretModel,Kempers2001_SoretModel,Mariana2003_SoretModelEval,Artola2008_SoretModel,Hoang2022_SoretModels} 
Especially in earlier works, the discrepancies can, at least in part, be attributed to inaccuracies in experimentally determined properties.
For example the Haase~\cite{Haase1950_SoretModel, Kempers2001_SoretModel} and Kempers~\cite{Kempers2001_SoretModel} models are very sensitive to partial molar properties, and therefore the equation of state used.\cite{Kempers2001_SoretModel,Mariana2003_SoretModelEval} 
In computer simulations all the required quantities can be accurately calculated, and as shown here using an exact model, all the theories examined herein feature noticeable deviations from the $S_T$ values obtained by direct simulation. 
Previous simulations have shown that the theories are accurate only in a very limited number of cases, even for simple LJ mixtures and hard-sphere mixtures.~\cite{Artola2008_SoretModel,Artola2007_SoretChemContr,Hoang2022_SoretModels}
We have tested the standard theoretical models against our simulation data. 
We calculate the Soret coefficient according to the models of Haase~\cite{Haase1950_SoretModel, Kempers2001_SoretModel} ($S_T^\mathrm{H}$),
Kempers~\cite{Kempers2001_SoretModel} ($S_T^\mathrm{K}$), 
Shukla and Firoozabadi~\cite{Shukla1998_SoretModel} ($S_T^\mathrm{SF}$), 
and Artola, Rousseau and Galli\'{e}ro~\cite{Artola2008_SoretModel} / Prigogine~\cite{Prigogine1950a_SoretModel, Prigogine1950b_SoretModel} ($S_T^\mathrm{ARG/P}$). Details about these models, along with additional results and analysis, are given in {sec.~1.4 of the SI}.

\begin{figure}[t]
    \centering
    \includegraphics[width=0.45\textwidth]{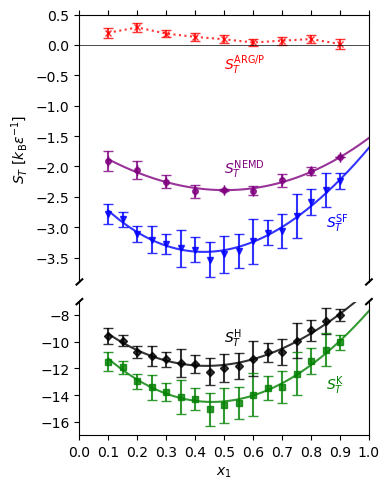}
    \caption{Soret coefficient $S_T$ as a function of $x_1$ as predicted by various models, as defined in the main text. The NEMD values $S_T^\mathrm{NEMD}$ are shown for reference. The solid lines show cubic functions fit to the $S_T(x_1)$ data.
    }
    \label{fig:STmodels}
\end{figure}

We show in fig.~\ref{fig:STmodels} the Soret coefficients predicted by these models, alongside the NEMD values for reference.
Out of the four models, the Shukla-Firoozabadi model is the most accurate: it overestimates $|S_T|$ by {$\sim$20-50\%}.
The Kempers and Haase models overestimate $|S_T|$ by {$\sim$400-500\%} and {$\sim$300-400\%} respectively. 
The Artola-Rousseau-Galli\'{e}ro/Prigogine model underestimates $|S_T|$ by {$\sim$100-110\%}, predicting values {$\sim 10^{-1}$~{$k_\mathrm{B} \epsilon^{-1}$}}. Furthermore, the model predicts the wrong sign: $S_T^\mathrm{ARG/P} > 0$ or straddles 0 when accounting for the associated uncertainties.

$S_T^\mathrm{H}$, $S_T^\mathrm{K}$ and $S_T^\mathrm{SF}$ all possess a minimum because they contain $x_1 (\partial \mu_1 / \partial x_1)_{P,T} = k_\mathrm{B} T \Gamma$ in the denominator (see the SI). The $x_1 (\partial \mu_1 / \partial x_1)_{P,T}$ term in $S_T^\mathrm{SF}$ originates directly from the phenomenological equations for thermodiffusion from linear non-equilibrium thermodynamics (LNET), which the model uses as a starting point for its derivation. For $S_T^\mathrm{K}$, the $x_1 (\partial \mu_1 / \partial x_1)_{P,T}$ term arises naturally from the statistical thermodynamics approach employed by Kempers. Originally an educated guess, $S_T^\mathrm{H}$ can be derived more rigorously within the framework of Kempers~\cite{Kempers2001_SoretModel}.   
In contrast, $S_T^\mathrm{ARG/P}$ does not predict a minimum: it features a weak concentration dependence, and generally decreases with increasing $x_1$. $S_T^\mathrm{ARG/P}$ contains $k_\mathrm{B} T$ (as opposed to $k_\mathrm{B}T \Gamma$) in the denominator, which is only valid for ideal mixtures. Indeed the Artola-Rousseau-Galli\'{e}ro/Prigogine model does not explicitly consider a concentration gradient along the reaction coordinate; doing so would result in a similar ``non-ideality'' term in the denominator.\cite{Rutherford1954_ThermalDiffusionModel}  

\textit{Closing remarks:}  
We close this letter with a discussion of our results in the context of recent literature.
While the $S_T$ minimum examined in this work originates from a coincident minimum in $\Gamma$, for
LiCl solutions, $\Gamma$ increases monotonically in the concentration range of the $S_T$ minimum.\cite{DiLecce2017_SoretLiClHeatOfTransport} 
Furthermore, simulations have highlighted the importance of ion solvation structure on the existence of the minimum in {LiCl$_\mathrm{(aq)}$}~\cite{DiLecce2017_SoretLiClIonWaterInteractions}. In contrast, local structural changes in the LJ mixture are minor {(see sec.~1.5 in the SI)}.  
We note that many simple salts are thought to have a $D_{12}$ minimum at low concentrations ($\sim 10^{-1}$~{mol dm$^{-3}$}) often attributed to ion-pair formation and solute-solvent association.~\cite{Gao2007_MutualDiffusionSalts,Mitchell1992_MutualDiffusionKSCN,Katz1980_DiffusionSeawater,Mohanakumar2021_ThermodiffusionAqueousSolutions}
Recent experiments observed $S_T$ minima in {NaSCN$_\mathrm{(aq)}$}, {KSCN$_\mathrm{(aq)}$}, {K$_2$CO$_{3\mathrm{(aq)}}$} and {CH$_3$COOK$_\mathrm{(aq)}$} that are carried through $D_T$.\cite{Mohanakumar2021_ThermodiffusionAqueousSolutions,Mohanakumar2022}
$S_T$ minima have been observed in mixtures of polar organic solvents with water. 
Ethanol/water and acetone/water mixtures have $S_T$ minima at $x_1 \sim 0.6$ and $x_1 \sim 0.5$, respectively, roughly coincident with minima in $D_{12}$ and $\Gamma$ for both mixtures.~\cite{Koniger2009_SoretMeasurements,Wiegand2004_SoretReview,Zhang2006_EthanolWater,Ning2006_SoretPolarSolvents,Cabrera2009_SoretAcetoneWater,Tyn1975_MutualDiffusionBinaryLiquidMixtures,Taylor1991_ActivityCoefDerivatives} 
Hence, the physical origin of the minima in aqueous solutions and mixtures might be quite different to the one reported here for non-polar liquids, since the LJ mixtures do not feature extrema in either $D_{12}$ or $D_T$, which instead change monotonically with composition (see fig.~\ref{fig:main}(b)).

LJ mixtures are more representative of mixtures of non-polar organic solvents, in which intermolecular interactions are governed by van der Waals forces.
At {25{\textdegree}C}, cyclohexane/cis-decalin possesses a $S_T$ maximum at {$x_1 \approx 0.2$} but $\Gamma \approx 1.0$ for the entire $0 \leq x_1 \leq 1$ range (there is a very shallow $\Gamma$ minimum at $x_1 \approx 0.8$), indicative of a different origin compared to the LJ mixture.   
Also in contrast with the LJ mixture, cyclohexane/benzene features a monotonic increase of $S_T$ with $x_1$ at {25{\textdegree}C}, but does have a strong minimum in $\Gamma$ at $x_1 \approx 0.5$.\cite{Hartmann2014_ThermophobicitySoret} 

It is evident that the physical origins of $S_T$ minima/maxima must be considered on a case-by-case basis for different mixtures, even those belonging to the same class of mixture (aqueous solution, non-polar organic solvent mixtures, etc.). We provide here a proof of principle and demonstrate that $S_T$ minima can exist in even the simplest of mixtures, as exemplified by a binary LJ mixture in which the components differ by only the interaction parameter $\epsilon$.     
The concentration dependence of the $S_T$ is, at least for simple mixtures, typically attributed to the cross-interactions between unlike particles\cite{Artola2007_SoretChemContr} -- a notion sustained in recent reviews\cite{Artola2013_ThermalDiffusionReview,Kohler2016_ThermalDiffusionReview,Harstad2009_ThermalDiffusionReview}. For dense supercritical LJ mixtures with cross-interactions given by $\epsilon_{12} = k_{12} \sqrt{\epsilon_{11} \epsilon_{22}}$ it was found that $S_T(x_1) \approx b x_1 + c$, with the slope $b$ controlled by $k_{12}$.\cite{Artola2007_SoretChemContr} Greater $|1 - k_{12}|$ values resulted in greater $|b|$ values.   
In this work, we identify a mixture with $k_{12} = 1$ that nevertheless features a strong composition dependence and more complex phenomenology (the $S_T$ minimum). Clearly, further work is required to explain the composition dependence of $S_T$ in liquid mixtures and different thermodynamic conditions.

\begin{acknowledgments}
We thank the Leverhulme Trust for Grant No. RPG-2018-384. 
We gratefully acknowledge a PhD studentship (Project Reference 2135626) for O.R.G. sponsored by ICL's Chemistry Doctoral Scholarship Award, funded by the EPSRC Doctoral Training Partnership Account (EP/N509486/1). 
We acknowledge the ICL RCS High Performance Computing facility and the UK Materials and Molecular Modelling Hub for computational resources, partially funded by the EPSRC (Grant Nos. EP/P020194/1
and EP/T022213/1).
\end{acknowledgments}

\bibliography{refs}

\end{document}


\maketitle

\newpage
\tableofcontents

\newpage

\section{Additional results \& discussion}

\subsection{Equation of state and phase diagram}

\begin{figure}[!b]
    \centering
    \begin{overpic}[width=0.9\textwidth,grid=False]{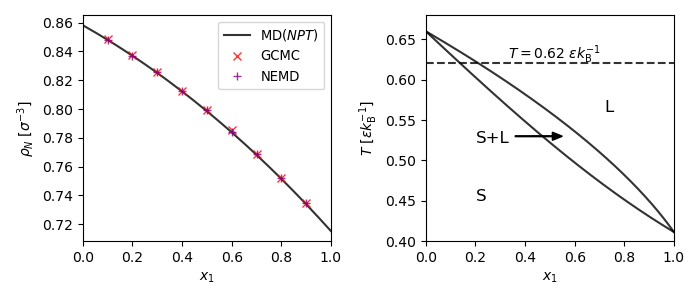}
        \put(0,40) {{\textsf{(a)}}}
        \put(52,40) {{\textsf{(b)}}}
    \end{overpic}
    \caption{Equation of state and phase diagram of the LJ mixture. (a) The number density $\rho_N$ for ({$P=0.46$~{$\epsilon \sigma^{-3}$}}, {$T=0.62$~{$\epsilon k_\mathrm{B}^{-1}$}}) as a function of mole fraction $x_1$, as predicted by MD(\textit{NPT}), GCMC and NEMD simulations. All simulated systems correspond to a liquid, which is metastable for $x_1 \leq x_l$, where $x_l$ is the mole fraction of the liquidus shown in (b).
    (b) The estimated solid solution phase diagram of the mixture at {$P=0.46$~{$\epsilon \sigma^{-3}$}}. The symbols are: S = solid solution; L = liquid mixture; S+L = solid and liquid coexistence.}
    \label{fig:EOSandPD}
\end{figure}

Fig.~\ref{fig:EOSandPD}(a) shows the equation of state of the mixture as a function of $x_1$. The equilibrium-NPT and NEMD results are in excellent agreement, indicating that the local equilibrium hypothesis is fulfilled in the NEMD simulations. 
The mixture is expected to have a solid-solution phase diagram,\cite{Hitchcock1999_LJSolidLiquidPD} and the thermodynamically stable phase for $x_1 = 0$ is a solid {(see sec.~\ref{subsec:melting_points} below)}. Therefore, the mixture crosses three regions of the phase diagram along the 
($x_1$; {$P=0.46$~{$\epsilon \sigma^{-3}$}}, {$T=0.62$~{$\epsilon k_\mathrm{B}^{-1}$}}) 
line: a liquid-liquid mixture for $x_l < x_1 < 1$ where $x_l$ corresponds to the liquidus; a liquid-solid coexistence region $x_s \leq x_1 \leq x_l$ where $x_s$ corresponds to the solidus; and a solid-solid mixture for $0 < x_1 < x_s$. All the results presented in this work correspond to a liquid-liquid mixture, including $x_1 \leq x_l$ for which our simulations are of the metastable supercooled liquid-liquid mixture. Thus, it is necessary to estimate $x_l$ in order to gauge the range of validity of our results.  

For an ideal solution the phase diagram can be estimated using the equations\cite{Galvin2021_binaryPD,Thurmond1953_ThermochemGeAlloys}
\begin{equation}\label{eq:PS1}
    \ln{ \frac{x_{s,1}}{x_{l,1}} } 
    = \frac{\Delta H_{\mathrm{fus},1}}{R} \left( \frac{1}{T_{m,1}} - \frac{1}{T} \right)
\end{equation}
\begin{equation}\label{eq:PS2}
    \ln{ \frac{1 - x_{s,1}}{1 - x_{l,1}} } 
    = \frac{\Delta H_{\mathrm{fus},2}}{R} \left( \frac{1}{T_{m,2}} - \frac{1}{T}  \right)
\end{equation}
where $R$ is the gas constant; $T_{m,i}$ and $\Delta H_{\mathrm{fus},i}$ are the melting temperature and enthalpy of fusion, respectively, of species $i$; and $x_{l,1}$ and $x_{s,1}$ are the mole fractions of the liquidus and solidus for end-member $1$. Eqs.~{\ref{eq:PS1}\&\ref{eq:PS2}} further assume that there is no difference in isobaric heat capacity between the supercooled liquid and pure solid of species $1$ at $T$.~\cite{Thurmond1953_ThermochemGeAlloys}
The predicted phase diagram is shown in fig.~{\ref{fig:EOSandPD}(b)}, with {$x_l \approx 0.21$} at {$T = 0.62$~{$\epsilon k_\mathrm{B}^{-1}$}}. 
The mixture is near-ideal at $x_1 \approx 0.21$ with thermodynamic factor $\Gamma \approx 0.8$, and the assumption of ideal mixing is therefore expected to be a good approximation.  
The $S_T$ and $\Gamma$ minima at $x_1 \sim 0.5$ (see the main text) are safely within the liquid-liquid mixture region of the phase diagram.

\subsection{Ideal and excess contributions to the thermodynamic factor}

\begin{figure*}[!h]
    \centering
    \includegraphics[width=0.5\textwidth]{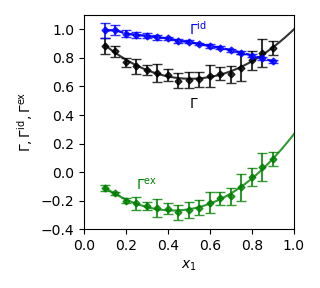}
    \caption{The thermodynamic factor $\Gamma$, and its ideal $\Gamma^\mathrm{id}$ and excess $\Gamma^\mathrm{ex}$ contributions, as a function of mole fraction $x_1$. All results correspond to the FEP data.
    }
    \label{fig:TFidex}
\end{figure*}

We split $\Gamma$ into its excess and ideal gas contributions, $\Gamma^\mathrm{ex}$ and $\Gamma^\mathrm{id}$ respectively (fig.~\ref{fig:TFidex}).
$\Gamma^\mathrm{id} = x_1 (\partial \mu_1^\mathrm{id} / \partial x_1)_{P,T} / (k_B T ) =(x_1/\rho_N) (\partial \rho_N / \partial x_1)_{P,T} + 1$ 
depends only on $x_1$, the total number density $\rho_N$ and its derivative, while 
$\Gamma^\mathrm{ex} = x_1 (\partial \mu_1^\mathrm{ex} / \partial x_1)_{P,T} / (k_B T )$ 
depends on the excess chemical potential $\mu_1^\mathrm{ex}$ and therefore the inter-particle interactions in the system. 
$\Gamma^\mathrm{id}$ monotonically decreases with increasing $x_1$, while $\Gamma^\mathrm{ex}$ possesses a minimum at {$x_1^{\mathrm{min}(\Gamma^\mathrm{ex})} = 0.4 \pm 0.1$} as determined by fitting a cubic function. Thus, the minimum arises through $\Gamma^\mathrm{ex}$, although its position is shifted by $\sim 0.1$ (to $x_1 \sim 0.5$) due to the $\Gamma^\mathrm{id}$ contribution.   

\subsection{Theoretical models}

\begin{figure*}[!t]
    \centering
    \begin{overpic}[width=0.98\textwidth,grid=False]{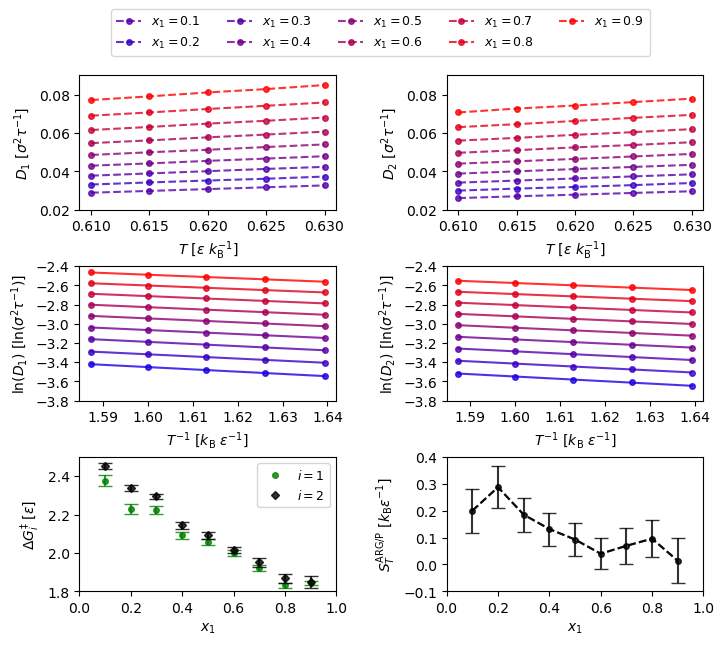}
        \put(0,78.5) {{\textsf{(a)}}}
        \put(50,78.5) {{\textsf{(b)}}}
        \put(0,53) {{\textsf{(c)}}}
        \put(50,53) {{\textsf{(d)}}}
        \put(0,26) {{\textsf{(e)}}}
        \put(50,26) {{\textsf{(f)}}}
    \end{overpic}
    \caption{The Soret coefficient predicted by the Artola-Rousseau-Galli\'{e}ro and Prigogine models, $S_T^\mathrm{ARG/P}$, and related quantities. The self-diffusion coefficients $D_i$ as a function of temperature $T$ for species (a) $i=1$ and (b) $i=2$. $\ln{D_i}$ \textit{vs.} $T^{-1}$ for species (c) $i=1$ and (d) $i=2$; the solid lines show linear fits to $\ln{D_i} = \ln{D_i^0} - \Delta G_i^{\ddag}/(k_B T)$. (e) The activation energies $\Delta G_i^{\ddag}$ and (e) $S_T^\mathrm{ARG/P}$ as a function of mole fraction $x_1$. Where error bars have not been shown, uncertainties are smaller than the size of the symbols.}
    \label{fig:STmodel_ARGP}
\end{figure*}

We calculate the Soret coefficient according to the models of Haase~\cite{Haase1950_SoretModel,Kempers2001_SoretModel} ($S_T^\mathrm{H}$), Kempers~\cite{Kempers2001_SoretModel} ($S_T^\mathrm{K}$) Shukla and Firoozabadi~\cite{Shukla1998_SoretModel} ($S_T^\mathrm{SF}$), and Artola, Rousseau and Galli\'{e}ro~\cite{Artola2008_SoretModel} ($S_T^\mathrm{ARG}$):
\begin{equation}\label{eq:ST_Haase}
    T S_{T,1}^\mathrm{H} = 
    \frac{m_1 m_2}{m_1 x_1 + m_2 x_2}
    \frac{(h_2 - h_2^0)/m_2 - (h_1 - h_1^0)/m_1}{x_1 (\partial \mu_1 / \partial x_1)_{P,T}}
    + \frac{RT^2}{x_1 (\partial \mu_1 / \partial x_1)_{P,T}} S_T^0
\end{equation}
\begin{equation}\label{eq:ST_Kempers}
    T S_{T,1}^\mathrm{K} = 
    \frac{v_1 v_2}{v_1 x_1 + v_2 x_2}
    \frac{(h_2 - h_2^0)/v_2 - (h_1 - h_1^0)/v_1}{x_1 (\partial \mu_1 / \partial x_1)_{P,T}}
    + \frac{RT^2}{x_1 (\partial \mu_1 / \partial x_1)_{P,T}} S_T^0
\end{equation}
\begin{equation}
    T S_{T,1}^\mathrm{SF} = 
    \frac{u_1/\tau_1 - u_2/\tau_2}{x_1 (\partial \mu_1 / \partial x_1)_{P,T}}
    + \frac{(v_2 - v_1) (x_1 u_1/\tau_1 + x_2 u_2/\tau_2) }{(x_1 v_1 + x_2 v_2) x_1 (\partial \mu_1 / \partial x_1)_{P,T}}
\end{equation}
\begin{equation}
    T S_{T,1}^\mathrm{ARG} = \frac{\Delta G_2^{\ddag} - \Delta G_1^{\ddag}}{RT} 
    + \frac{m_2 - m_1}{m_2 + m_1} \frac{\Delta G_2^{\ddag} + \Delta G_1^{\ddag}}{RT} 
\end{equation}
where $h_i$, $u_i$ and $v_i$ are the partial molar enthalpy, internal energy and volume of species $i=1,2$. $R$ is the gas constant.
The Kempers model was derived by considering a non-isothermal two-bulb system, with the main assumption that the stationary state is the macroscopic state with the maximum number of microstates. Within this model, $S_T^\mathrm{K}$ corresponds to the centre-of-volume frame of reference, while Haase's earlier educated guess $S_T^\mathrm{H}$ can be derived in the centre-of-mass frame.~\cite{Kempers2001_SoretModel} $h_i^0$ and $S_T^0$ correspond to an ideal gas state at the same temperature (calculated from kinetic theory), and capture the kinetic contribution to $S_T$. 
The Shukla-Firoozabadi model was developed with its origins in linear non-equilibrium thermodynamics (LNET), along the same lines as earlier models\cite{Dougherty1955_ThermalDiffusionJCP,Dougherty1955_ThermalDiffusionJPC} that all correlate the net heat of transport with the activation energy for viscous flow $\Delta U_\eta^\ddag$. The parameter $\tau_i = \Delta U_{c,i}/ \Delta U_{\eta,i}^\ddag$, where $\Delta U_c$ is the cohesive energy, is related to the size of the hole required for viscous flow, and is often treated as an adjustable parameter. 
The Artola-Rousseau-Galli\'{e}ro model is Prigogine's model ($S_T^\mathrm{P}$) modified to include the mass contribution; for $m_1 = m_2$ it reduces to $S_{T,1}^\mathrm{ARG} = S_{T,1}^\mathrm{P} = (\Delta G_2^\ddag - \Delta G_1^\ddag)/RT^2$. In these two kinetic models, thermal diffusion is described as a coupled diffusion-activated process, for which the elementary process can be summarized as a swap between two particles of different species along a temperature gradient. The activation energies $\Delta G_i^\ddag$ were calculated from $D_i = D_i^0 \exp{(-\Delta G_i^\ddag / RT)}$ (see {fig.~\ref{fig:STmodel_ARGP}}), where $D_i^0$ is a constant and the self-diffusion coefficients $D_i$ have been corrected for finite-size effects (see sec.~\ref{subsec:diff_visc} below).  
The activation energy transported in molecular motion is better identified with self-diffusion than viscous flow; the use of the latter was historically motivated by the scarcity of self-diffusion data, even for pure components, and justified by the similar activation energies expected from Eyring's rate theory applied to liquids.\cite{Tichacek1956_SoretModel}   
By using self-diffusion data, the Artola-Rousseau-Galli\'{e}ro model represents a step forward in the modelling of thermal diffusion.

\begin{figure*}[t]
    \centering
    \begin{overpic}[width=0.9\textwidth,grid=False]{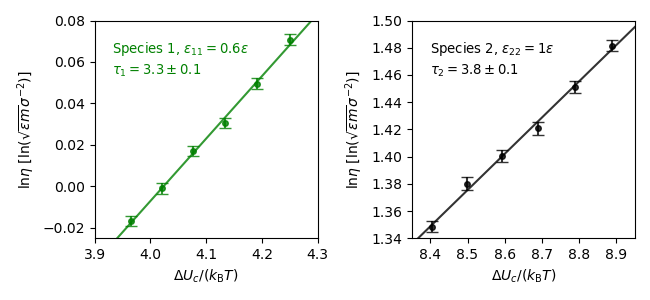}
        \put(0,43.8) {{\textsf{(a)}}}
        \put(51,43.8) {{\textsf{(b)}}}
    \end{overpic}
    \caption{$\ln(\eta_i)$ \textit{vs.} $\Delta U_c / (k_B T)$, where $\eta_i$ and $\Delta U_c$ are the shear viscosity and cohesive energy of a pure liquid of species $i=1,2$. (a) Species 1 and (b) species 2. The solid lines show linear fits to $\ln{\eta_i} = \ln{A} + U_c/(\tau_i k_B T)$. The data points correspond to $P = 0.46 \epsilon \sigma^{-3}$ and $T/(\epsilon k_B^{-1}) =0.605$, $0.61$, $0.615$, $0.62$, $0.625$ and $0.63$. }
    \label{fig:viscfit}
\end{figure*}

A relevant question for the Shukla-Firoozabadi model is how to best estimate the parameters $\tau_i$. Different methods have been proposed, with varying levels of approximations.\cite{Shukla1998_SoretModel,Eslamian2009_SoretModelReview,Eslamian2009_DynamicTDmodel,Yan2008_TDmodelsHydrocarbons,Pan2007_TDmodelsAlkanolSolutions} The most crude but still a widely 
{adopted}~\cite{Artola2007_SoretChemContr,Hoang2022_SoretModels} 
approach is to set $\tau_1 = \tau_2 = 4.0$ or 3.5, stemming from the observation~\cite{Shukla1998_SoretModel} that $\Delta U_\mathrm{vap}/\Delta U_{\eta}^\ddag = 3$-$4$ for many non-associating liquids under normal boiling point conditions, where $\Delta U_\mathrm{vap}$ is the energy of vaporization ($\Delta U_\mathrm{vap}$ is an approximation to $\Delta U_c$). While permissible for some mixtures, this approach is unsuitable for others and has been criticised.\cite{Eslamian2009_SoretModelReview,Pan2007_TDmodelsAlkanolSolutions,Yan2008_TDmodelsHydrocarbons} 
In this work, we use the method originally proposed~\cite{Shukla1998_SoretModel}, but not employed, by Shukla and Firoozabadi with the caveat that they used the approximation $\Delta U_\mathrm{vap} \approx \Delta U_c$. 
As shown in {fig.~\ref{fig:viscfit}}, we calculate $\tau_i$ of the pure components from $\eta_i = A \exp{(\Delta U_{\eta,i}^\ddag/RT)} = A \exp{(\Delta U_{c,i} / \tau_i R T)}$, where $\eta_i$ is the shear viscosity, to give {$(\tau_1,\tau_2) = (3.3 \pm 0.1, 3.8 \pm 0.1)$}, and use these $\tau_i$ for the entire composition range. This approach does not account for the composition dependence of $\tau_i$.

\newpage
\subsection{Radial distribution functions}

\begin{figure*}[!h]
    \centering
    \begin{overpic}[width=1.0\textwidth,grid=False]{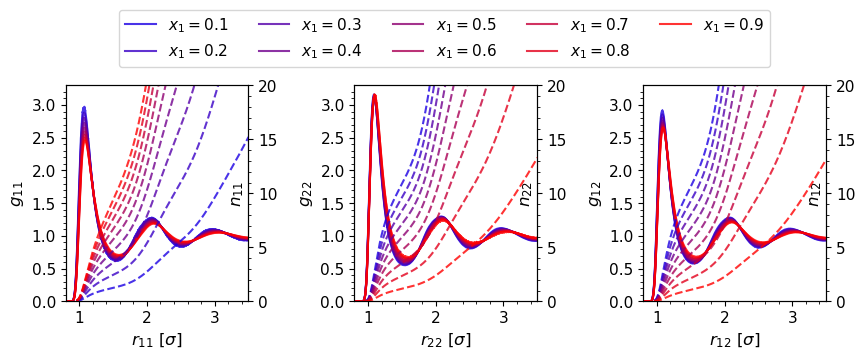}
        \put(24, 9) {{\textsf{(a)}}}
        \put(57.5, 9) {{\textsf{(b)}}}
        \put(91, 9) {{\textsf{(c)}}}
    \end{overpic}
    \caption{Radial distribution functions $g_{ij}$ and coordination number $n_{ij}$ of the Lennard-Jones mixtures: (a) $g_{11}$ and $n_{11}$; (b) $g_{22}$ and $n_{22}$; (c) $g_{12}$ and $n_{12}$. $r_{ij}$ is the radial distance between species $i$ and $j$. Solid lines show $g_{ij}$ (left axis) and the dashed lines show $n_{ij}$ (right axis).}
    \label{fig:RDFs}
\end{figure*}

\subsection{Thermal conductivity}

It is well established that the Soret effect reduces the thermal conductivity of the mixture.\cite{GrootMazurNET} 
From LNET, the thermal conductivity $\lambda$ is given in terms of the phenomenological coefficients $L_{\alpha \beta}$ and $L_{\alpha \beta}^\prime$ as 
\begin{equation}
    \lambda = \frac{1}{T^2} \left( L_{qq} - \frac{L_{1q}L_{q1}}{L_{11}} \right)
    = \frac{1}{T^2}  \left( L_{qq}^{\prime} - \frac{L_{1q}^{\prime}L_{q1}^{\prime}}{L_{11}}  \right)
\end{equation}
where the enthalpic terms in the primed coefficients cancel exactly to give the same thermal conductivity as for the unprimed coefficients. $\lambda$ can therefore be split into two contributions: (1) the thermal conductivity in the absence of coupling effects,
$\lambda_0 = L_{qq}/T^2$ or $\lambda_0^\prime = L_{qq}^\prime/T^2$, 
and (2) the mass-heat coupling term, 
$\delta \lambda = -{L_{1q}L_{q1}}/(L_{11} T^2)$ or 
$\delta \lambda^\prime = -{L_{1q}^{\prime} L_{q1}^{\prime}}/(L_{11} T^2)$, 
such that $\lambda = \lambda_0 + \delta \lambda = \lambda_0^\prime + \delta \lambda^\prime$.
The Soret effect should always decrease $\lambda$, and thus $\lambda = \lambda_0 - |\delta \lambda| = \lambda_0^\prime - |\delta \lambda^\prime|$.

\begin{figure*}[!t]
    \centering
    \begin{overpic}[width=1.0\textwidth,grid=False]{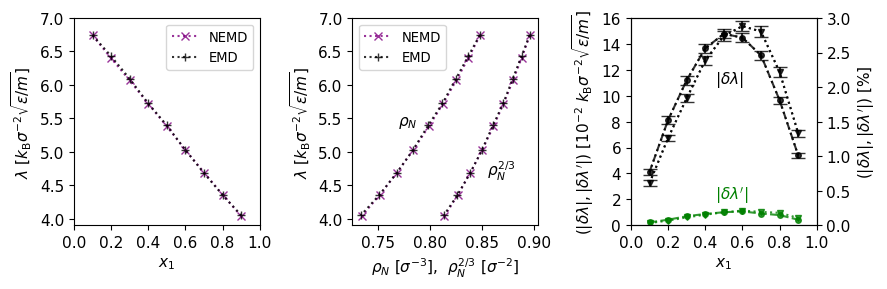}
        \put(0, 32) {{\textsf{(a)}}}
        \put(31, 32) {{\textsf{(b)}}}
        \put(61, 32) {{\textsf{(c)}}}
    \end{overpic}
    \caption{Thermal conductivity $\lambda$ of the Lennard-Jones mixture. $\lambda$ as a function of (a) mole fraction $x_1$ and (b) number density $\rho_N$ as well as $\rho_N^{2/3}$. (c) The decrease in $\lambda$ due to heat-mass coupling, $|\delta \lambda|$ and $\delta \lambda^\prime$. In (c), circles with dashed lines (-{}-$\circ$-{}-) show the absolute values (left axis) and downward triangles with dotted lines ($\cdot \cdot \triangledown \cdot \cdot$) show the values as a percentage of $\lambda$ (right axis).
    Where error bars have not been shown, uncertainties are smaller than the size of the symbols.}
    \label{fig:TC}
\end{figure*}

We show in fig.~\ref{fig:TC} the thermal conductivity of the mixture, and the contribution due to heat-mass coupling. The thermal conductivities calculated via NEMD and equilibrium-MD (EMD) are in excellent agreement. The thermal conductivity of the mixture decreases with increasing $x_1$. $\lambda$ increases with density, consistent with the trend observed for simple fluids, including LJ fluids, along an isotherm. However, the density dependence of the mixture slightly deviates from the $\lambda \propto \rho_N^{2/3}$ scaling reported\cite{Ohtori2014_TCsimpleliq} for pure LJ fluids. 
The $|\delta \lambda|$ and $|\delta \lambda^\prime|$ terms both feature a maximum, which is a direct consequence of the minimum in $S_T$ and $\Gamma$ 
{(indeed, $\delta \lambda^\prime = - L_{1q}^\prime S_T \Gamma k_\mathrm{B}/m_1$ for $m_1 = m_2$)}.
$|\delta \lambda|$ amounts to a reduction of {$\sim$1-3\%} relative to $\lambda$ for $0.1 \leq x_1 \leq 0.9$, while $|\delta \lambda^\prime|$ is an order of magnitude smaller with values {from 0 to 0.3\%}. The result that heat-mass coupling reduces the thermal conductivity by at most a few percent is consistent with previous studies\cite{Armstrong2014_TCsoret} examining different LJ mixtures.

\newpage
\section{Physical properties from simulations}

All simulations were performed using the software package LAMMPS~\cite{LAMMPS} (v. 3 March 2020).

\subsection{Equilibrium molecular dynamics simulations}

Equilibrium molecular dynamics (EMD) simulations of the mixture were performed in the \textit{NPT}, \textit{NVT} and \textit{NVE} ensembles, targeting various thermodynamic states. Unless stated otherwise (i.e. for the finite-size analyses) a cubic simulation cell containing 5000 particles was used. A timestep of {$0.002 \tau$} was used for the \textit{NPT} and \textit{NVT} simulations.  
For the \textit{NPT} and \textit{NVT} simulations, temperature was controlled by the Nos\'{e}-Hoover chain thermostat, with 3 chains, and a time constant of {1~$\tau$}. Additionally in the \textit{NPT} simulations, pressure was controlled using a Nos\'{e}-Hoover chain barostat, also with 3 chains, and a time constant of {4~$\tau$}. For the \textit{NPT} simulations, a single replica was performed for each ($x_1$, $P$, $T$) state point, consisting of at least {$2 \times 10^4$~{$\tau$}} of equilibration, followed by a {1-2$\times 10^5$~{$\tau$}} production run. For the \textit{NVT} simulations, sampling consisted of 20-200 replicas for each ($x_1$, $\rho$, $T$) state point depending on the system size, which varied from {$N=1000$-$120,000$} particles; each replica was equilibrated for {$2 \times 10^3$~{$\tau$}} followed by {$10^4$~{$\tau$}} of production. The exceptions to this were the large {$N = 0.5$-$5\times 10^6$} systems simulated for the KBIs ({sec.~\ref{sec:KBI_RDF} below}), which had both equilibration and production times of $10^3 \tau$.

Simulations in the \textit{NVE} ensemble targeted different compositions along the ({$P=0.46$~{$\epsilon \sigma^{-3}$}}, {$T = 0.62$~{$\epsilon k_\mathrm{B}^{-1}$}}) isobar-isotherm. For each state point, replicas were first spawned from \textit{NVT} simulations and monitored for an initial $2 \times 10^3$~{$\tau$}. If the average temperature was within {$\pm 0.0005$~{$\epsilon k_\mathrm{B}^{-1}$}} of {$T = 0.62$~{$\epsilon k_\mathrm{B}^{-1}$}}, the replica was continued for a further {$2 \times 10^4$~{$\tau$}} of production. Replicas were also subsequently started from these successful \textit{NVE} trajectories. A total of {50-100} statistically independent replicas were performed for each state point, all of which had an average temperature $\langle T \rangle_{NVE}$ within {$\pm 0.0005$~{$\epsilon k_\mathrm{B}^{-1}$}} of the target temperature.  
A smaller timestep of $\delta t = 0.001$~{$\tau$} was used to improve energy conservation (reduce temperature drift) and therefore increase the sampling times available in the small {$\pm 0.0005$~{$\epsilon k_\mathrm{B}^{-1}$}} temperature window, and also to reduce the discretization error for the numerical integration of correlation functions (see secs.~\ref{subsec:phemcoeff} \& \ref{subsec:diff_visc} below).

\subsection{Soret coefficient and thermal conductivity from NEMD}

\subsubsection{Simulation details}

\begin{figure*}[!b]
    \centering
    \begin{overpic}[width=0.8\textwidth,grid=False]{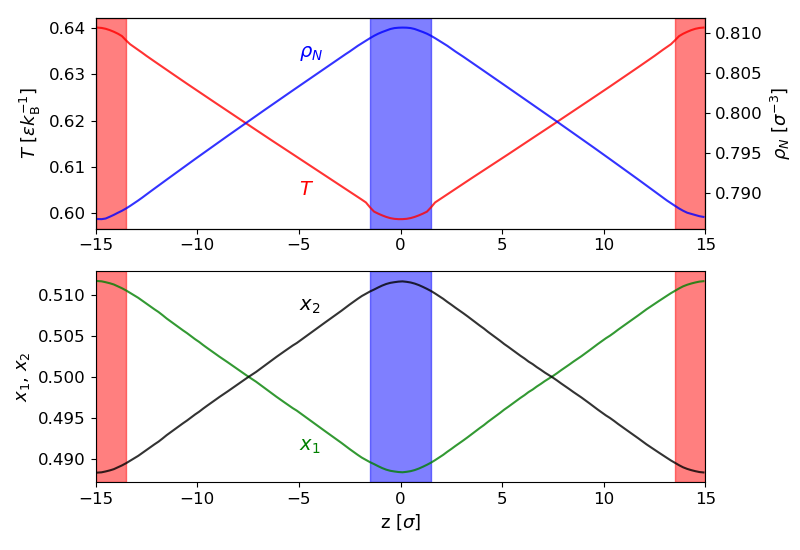}
        \put(0, 67) {{\textsf{(a)}}}
        \put(0, 35) {{\textsf{(b)}}}
    \end{overpic}
    \caption{Representative (a) temperature $T$ and number density $\rho_N$ profiles, and (b) mole fraction, $x_1$ and $x_2$, profiles for the NEMD simulations. The blue (cold) and red (hot) indicate the location of the thermostatting regions in the simulation cell.}
    \label{fig:nemd_profiles}
\end{figure*}

The Soret coefficient $S_T$ and thermal conductivity $\lambda$ were calculated from boundary-driven non-equilibrium molecular dynamics simulations (NEMD) in the stationary state. An elongated (tetragonal) simulation cell of dimensions $(L_x, L_y, L_z) = (20, 20, 30)\sigma$ was used, with 3D periodic boundary conditions. Two thermostatting regions, hot and cold, were located in the centre and edges of the simulation, respectively (see fig. \ref{fig:nemd_profiles}). The thermostatting regions had a width $\Delta z = 3\sigma$ and extended over the entire $(x,y)$ plane, such that the temperature gradients were generated along the $z$-direction. 
For the thermostatting procedure, a simple velocity rescaling algorithm was used to maintain the hot and cold thermostatting regions at temperatures $T_h$ and $T_c$, respectively.
The velocities of all particles in each region were rescaled, every timestep, by a factor $\alpha = \sqrt{K_t / K_c}$ where $K_t$ and $K_c$ are the target and current kinetic energies of the region. 
This velocity rescaling procedure does not conserve linear momentum, so the system’s centre-of-mass velocity was
subtracted from each particle at every time step in order to ensure linear momentum conservation. A timestep of {$\delta t = 0.002$~{$\tau$}} was used.

For each system, the average density, mole fraction and thermostat temperatures were chosen to give $\langle P_{zz} \rangle$ within {$\pm 0.001$~{$\epsilon \sigma^{-3}$}} of the target pressure {$P=0.46$~{$\epsilon \sigma^{-3}$}}, where $P_{zz}$ is the pressure tensor component parallel to the heat flux vector. Relatively small (by simulation standards) temperature differences {$\Delta T = T_h - T_c \approx 0.04$~{$\epsilon k_{B}^{-1}$}} and resulting gradients {$\nabla T \approx 2.9 \times 10^{-3}$~{$\epsilon k_{B}^{-1} \sigma^{-1}$}} were used to accurately target the thermodynamic state. 
For each system, 10 statistically independent replicas were generated, each consisting of an initial {$2 \times 10^4 \tau$} to establish the stationary state, followed by {$2 \times 10^6 \tau$} for data collection.

In the stationary state, this set-up results in two equal but opposite temperature gradients, and therefore in equal and opposite heat fluxes, such that the system is completely periodic. 
The heat flux across the system, $\bm{J}_q = (0,0,\pm J_q)$, can be obtained from the continuity equation
\begin{equation}
    J_q = \frac{|\langle \Delta U \rangle | }{2 \delta t A}
\end{equation}
where $A = L_x \times L_y$ is the cross-sectional area of the simulation cell, $\delta t$ is the timestep, and $\Delta U$ is the internal energy exchanged at each timestep. The employed simple rescaling thermostat only changes the kinetic energy such that $\Delta U = \Delta K$. The factor of 2 in the denominator accounts for the two heat fluxes (equal magnitude and opposite direction) generated in this setup. The thermal conductivity $\lambda$ was then calculated using Fourier's Law:
\begin{equation}
    \bm{J}_q = - \lambda \nabla T
\end{equation}\label{eq:Fourier_Law} 
where $\nabla T$ is the \textit{local} temperature gradient. The Soret coefficient $S_T$ was calculated using {eq.~1 in the main text}, again employing local values of the gradients $\nabla T$ and $\nabla x_1$.     
Local densities $\rho$, mol fractions $x_1$ and temperatures $T$ were determined from a {$1\sigma$} bin close to the centre of each \textit{NVE} compartment. The position of the bin was chosen such that {$T=0.62$~{$\epsilon k_\mathrm{B}^{-1}$}}. The local values of $\nabla T$ and $\nabla w_1$ were determined by fitting straight lines to the temperature and mole fraction profiles within a range of {$\pm 2.5 \sigma$} around the selected state point.

\subsubsection{Linear response}

\begin{figure*}[!t]
    \centering
    \begin{overpic}[width=0.95\textwidth,grid=False]{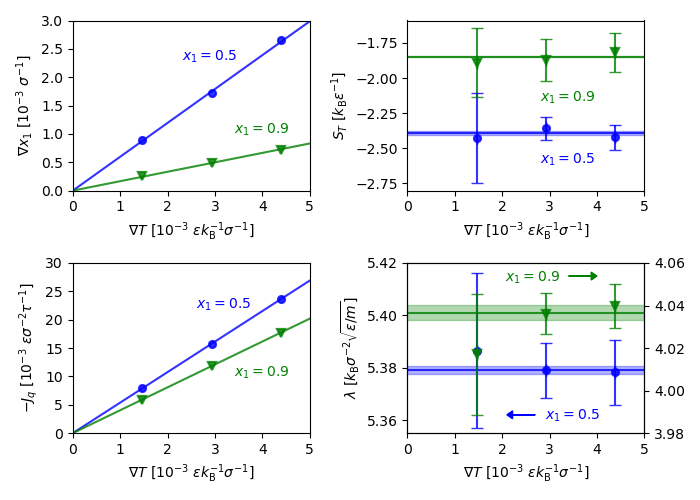}
        \put(0, 69) {{\textsf{(a)}}}
        \put(46, 69) {{\textsf{(b)}}}
        \put(0, 35) {{\textsf{(c)}}}
        \put(46, 35) {{\textsf{(d)}}}
    \end{overpic}
    \caption{Linear response of the Soret coefficient $S_T$ and thermal conductivity $\lambda$: (a) mole fraction gradient $\nabla x_1$, (b) $S_T$, (c) heat flux $J_q$ and (d) $\lambda$ as a function of temperature gradient $\nabla T$. The solid lines show fits to $\nabla x_1 = -x_1 x_2 S_T \nabla T$ in (a) and (b), and to Fourier's law $- J_q = \lambda \nabla T$ in (c) and (d). In (a) and (c), statistical uncertainties are smaller than the size of the symbols. In (b) and (d) the shaded areas show the uncertainty associated with the fits.
    }
    \label{fig:nemd_LR}
\end{figure*}

We demonstrate in {fig.~\ref{fig:nemd_LR}} that the magnitude of temperature gradients used in this work are within the linear regime.  
For these additional simulations, production runs of length $1.5 \times 10^{6}\tau$ and $10^{6}\tau$ were used for the 
{$\nabla T \approx 1.5 \times 10^{-3}$~{$\epsilon k_{B}^{-1} \sigma^{-1}$}}
and 
{$\nabla T \approx 4.4 \times 10^{-3}$~{$\epsilon k_{B}^{-1} \sigma^{-1}$}}
systems, respectively. 
The values for $S_T$ and $\lambda$ for $x_1 = 0.5,0.9$ were calculated by fitting to their linear response (fig.~\ref{fig:nemd_LR}).

\subsubsection{Finite-size effects}

\begin{figure*}[!h]
    \centering
    \includegraphics[width=0.5\textwidth]{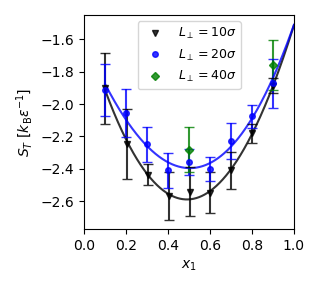}
    \caption{Finite-size effects on the Soret coefficient $S_T$ calculated from NEMD simulations. $L_\perp = L_x = L_y$ is the length of the simulation cell in the direction perpendicular to the heat flux. Solid lines show cubic functions fit to $S_T(x_1)$ where $x_1$ is the mole fraction of species 1.  
    }
    \label{fig:SoretFS}
\end{figure*}

We show in {fig.~\ref{fig:SoretFS}} that the lateral simulation cell length $L_\perp = L_x = L_y$ has a significant effect on $S_T$. While all but one ($L_\perp /\sigma = 10,20$ for $x_1 = 0.3$) $S_T$ values agree to within their statistical uncertainties, $|S_T|$ systematically decreases with $L_\perp$: by {$\sim$0-10\%} ({$\sim$3-6\%}) when increasing $L_\perp$ from $10\sigma$ to $20\sigma$ ($20\sigma$ to $40\sigma$). Nevertheless, fitting cubic functions to the $L_\perp / \sigma = 10, 20$ data gives $x_1^{\mathrm{min}(S_T)} = 0.5 \pm 0.1$, and further increasing $L_\perp$ is not expected to significantly shift the position of the minimum. Finite-size effects in $S_T$ are expected from those observed in self-diffusion coefficients (sec.~\ref{subsec:diff_visc} below) and $D_{12}$.~\cite{Jamali2018_D12FiniteSize,Celebi2021_diffFSrev} The impact of finite-size effects on $D_T$ is less well known.

We do not observe appreciable finite-size effects for the thermal conductivity. This is consistent with the mechanism of thermal transport in molecular liquids (and liquid mixtures), which is dominated by collisions between
nearest neighbors, setting a characteristic length scale for heat
transport at $\sim 1 \sigma$.

The additional simulations for $L_\perp = 10 \sigma$ and $L_\perp = 40 \sigma$ had production lengths of $4 \times 10^6 \tau$ and $3 \times 10^5 \tau$, respectively.

\subsection{Partial molar properties}

Partial molar properties $z_{i}$ were calculated using the equations $z_{1} = Z + (1-x_1)(\partial Z/\partial x_1)_{PTN_{2}}$ and $z_{2} = Z - x_1(\partial Z/\partial x_1)_{PTN_{2}}$, where $Z$ is the corresponding extensive property. $Z(x_1)$ were calculated from MD simulations in the $NPT$ ensemble. At each selected $x_1$, two additional simulations were performed at $x_1 \pm 0.01$, and $(\partial Z/\partial x_1)_{PTN_{2}}$ was then calculated by fitting a straight line through these three points.

\begin{figure*}[!h]
    \centering
    \begin{overpic}[width=1.0\textwidth,grid=False]{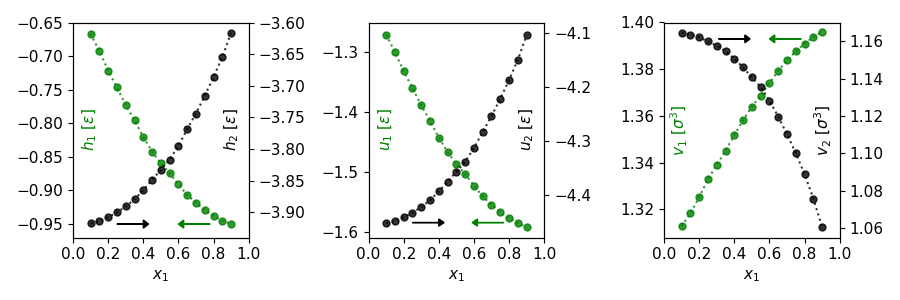}
        \put(0, 33) {{\textsf{(a)}}}
        \put(35, 33) {{\textsf{(b)}}}
        \put(65, 33) {{\textsf{(c)}}}
    \end{overpic}
    \caption{Partial molar properties $z_{i}$ of component $i=1,2$ as a function of mole fraction $x_1$: (a) partial molar enthalpy $h_{i}$; (b) partial molar internal energy $u_{i}$; and (c) partial molar volume $v_{i}$.
    Statistical uncertainties are smaller or comparable to the size of the symbols.} 
    \label{fig:partial_molar}
\end{figure*}

\newpage
\subsection{Onsager's phenomenological coefficients}\label{subsec:phemcoeff}

The phenomenological coefficients $L_{\alpha \beta}$ were calculated using the Green-Kubo integral formula
\begin{equation}\label{eq:GKintegral}
    L_{\alpha \beta} = \frac{V}{3 k_\mathrm{B}} \lim_{t^\prime \to \infty} \int_0^{t^\prime} \langle {\bm{J}_{\alpha}(t) \cdot \bm{J}_{\beta}(0)} \rangle dt
\end{equation}
where $V$ is the volume of the simulation cell, and the factor of 3 in the denominator averages the contributions from each spatial dimension. In order to calculate $L_{qq}$, $L_{1q}$, $L_{q1}$ and $L_{11}$, the expressions for the heat flux $\bm{J}_q$ and mass flux $\bm{J}_1$ in terms of microscopic quantities are required. These are:
\begin{equation}
    \bm{J}_1 = \frac{1}{V} \sum_{i=1}^{N_1} m_i \bm{v}_i
\end{equation}
where the sum runs over all $N_1$ particles of species 1, and in the case of two-body interactions the Irving-Kirkwood formula for heat flux is
\begin{equation}
    \bm{J}_q = \frac{1}{V} \left( \sum_{i=1}^N U_i \bm{v}_i - \sum_{i=1}^N \bm{S}_i \bm{v}_i \right)
     = \frac{1}{V} \left( 
    { 
    \sum_{i=1}^N (\mathcal{V}_i + K_i) \bm{v}_i 
    - \frac{1}{2} \sum_{i=1}^N \sum_{j \neq i}^N (\bm{v}_i \cdot \bm{F}_{ij}) \bm{r}_{ij}
    }
    \right)
\end{equation}
where $U_i$ is the per-particle internal energy which can be split into potential $\mathcal{V}_i = (1/2) \sum_{j \neq i} u_{ij}(r_{ij})$ and kinetic energy $K_i = (1/2) m_i \bm{v}_i^2$ contributions; $\bm{S}_i$ is the per-particle stress tensor; $\bm{F}_{ij}$ is the force exerted on particle $i$ by particle $j$; and $\bm{r}_{ij} = \bm{r}_j - \bm{r}_i$ where $\bm{r}_i$ is the position vector of particle $i$. 

\begin{figure}[!t]
    \centering
    \begin{overpic}[width=1.0\textwidth,grid=False]{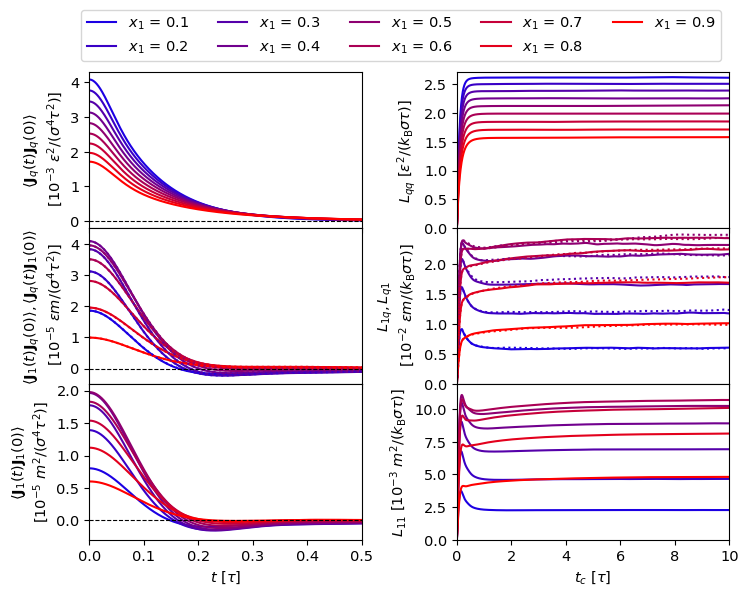}
        \put(0, 69) {\textbf{\textsf{(a)}}}
        \put(51, 69) {\textbf{\textsf{(b)}}}
        
        \put(43, 67) {{\textsf{(i)}}}
        \put(43, 46) {{\textsf{(ii)}}}
        \put(43, 25) {{\textsf{(iii)}}}
        
        \put(92, 51) {{\textsf{(i)}}}
        \put(92, 30) {{\textsf{(ii)}}}
        \put(92, 8.5) {{\textsf{(iii)}}}
    \end{overpic}
    \caption{ (a) Correlation functions $\langle \bm{J}_{\alpha}(t) \cdot \bm{J}_{\beta}(0) \rangle$ as a function of time lag $t$. 
    (b) Convergence of the phenomenological coefficients $L_{\alpha \beta}(t_c)$ with the correlation time $t_c$ used as the upper limit of the integral in {eq.~\ref{eq:GKintegral}}. In (a)(ii) and (b)(ii): solid lines show $\langle \bm{J}_{1}(t) \cdot \bm{J}_{q}(0) \rangle$ and $L_{1q}$; dotted lines show $\langle \bm{J}_{q}(t) \cdot \bm{J}_{1}(0) \rangle$ and $L_{q1}$
    }
    \label{fig:GK_CF_integral}
\end{figure}

The unprimed coefficients were calculated from simulations in the \textit{NVE} ensemble. A correlation time of $t_c = 10 \tau$ was used for the upper limit of {eq.~\ref{eq:GKintegral}}. Selecting this integration limit is a compromise between sampling efficiency and minimizing the resulting truncation error. We show in {fig.~{\ref{fig:GK_CF_integral}}} that 10~{$\tau$} is sufficiently long to achieve a well-converged integral, while the exhaustive extent of our sampling is reflected in the associated uncertainty.

The primed coefficients $L_{\alpha \beta}^{\prime}$ were then obtained from $L_{\alpha \beta}$ using the formulas
\begin{equation}
    L_{qq}^{\prime} = L_{qq} - (L_{1q}+L_{q1})(h_{s,1} - h_{s,2}) + L_{11}(h_{s,1} - h_{s,2})^2 
\end{equation}
\begin{equation}
    L_{1q}^{\prime} = L_{1q} - L_{11}(h_{s,1} - h_{s,2})
\end{equation}
\begin{equation}
    L_{q1}^{\prime} = L_{q1} - L_{11}(h_{s,1} - h_{s,2})
\end{equation}
\begin{equation}
    L_{11}^{\prime} = L_{11}
\end{equation}
where $h_{s,i}$ is the specific molar enthalpy of component $i=1,2$. These formulas can be derived by comparing expressions for entropy production in LNET, or trivially from the Green-Kubo integral formula (eq.~\ref{eq:GKintegral}) noting that $\bm{J}_q^{\prime} = \bm{J}_q - (h_{s,1} - h_{s,2}) \bm{J}_1$.~\cite{Miller2013_SoretArKr, Armstrong2014_TCsoret} 
The difference between $\bm{J}_q^{\prime}$ and $\bm{J}_q$ corresponds to the heat transported due to diffusion.

\begin{figure}[!h]
    \centering
    \begin{overpic}[width=1.0\textwidth,grid=False]{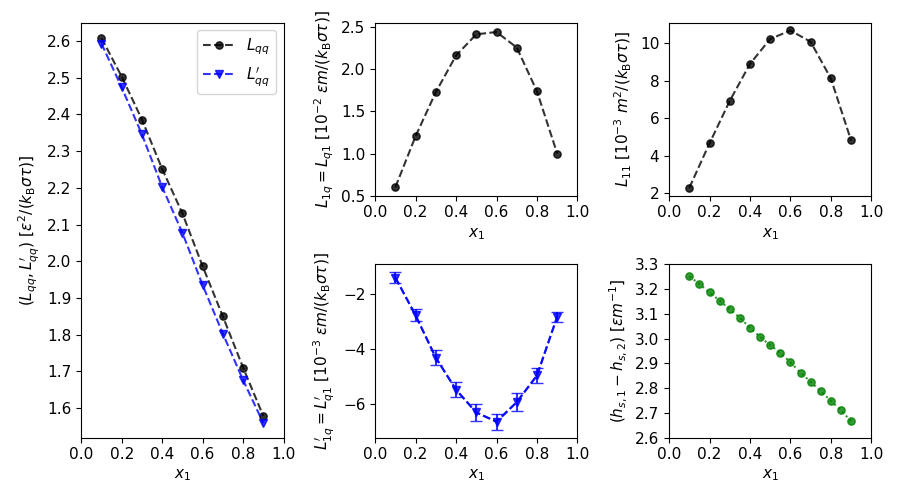}
        \put(10, 8.5) {{\textsf{(a)}}}
        \put(42.5, 50) {{\textsf{(b)}}}
        \put(75.5, 50) {{\textsf{(c)}}}
        \put(42.5, 8.5) {{\textsf{(d)}}}
        \put(75.5, 8.5) {{\textsf{(e)}}}
    \end{overpic}
    \caption{The phenomenological coefficients as a function of the mole fraction $x_1$: (a) $L_{qq}$ and $L_{qq}^{\prime}$; (b) $L_{1q} = L_{q1}$; (c) $L_{11}$; and (d) $L_{1q}^{\prime} = L_{q1}^{\prime}$. 
    (e) The difference in specific partial enthalpies $h_{s,1} - h_{s,2}$, where $h_{s,i}$ is the specific partial enthalpy of component $i$. Where error bars are not shown, the statistical uncertainty is smaller than the size of the symbol.}
    \label{fig:phemcoeff}
\end{figure}

We show in {fig.~\ref{fig:phemcoeff}} the values of coefficients $L_{\alpha \beta}$ and $L_{\alpha \beta}^{\prime}$. Consistent with Onsager's reciprocal relations, $L_{1q}^{\prime} = L_{q1}^{\prime}$ and $L_{1q} = L_{q1}$ to within their associated uncertainties. We therefore average over both reciprocal coefficients to give a single value each for $L_{1q}^{\prime} = L_{q1}^{\prime}$ and $L_{1q} = L_{q1}$.

\subsection{Chemical potentials}\label{subsec:FEP}

The chemical potential of species $i$ was calculated in terms of its ideal (id) and excess (ex) contributions, $\mu_i = \mu_i^{\mathrm{id}} + \mu_i^{\mathrm{ex}}$.
The ideal gas contribution $\mu^\mathrm{id}_i$ is given by the formula
\begin{equation}\label{FEid}
    \mu^{\mathrm{id}}_i = k_B T \ln{( \rho_{N,i} \Lambda_i^3 )}
\end{equation}
where $\rho_{N,i} = x_i \rho_N$ and $\Lambda_i$ are the number density and thermal de Broglie wavelength of species $i$, respectively. The densities were obtained from MD simulations in the $NPT$ ensemble. 
By convention we set ${\Lambda_2 =  h / \sqrt{ 2 \pi m_2 k_B T} = 1} \Rightarrow {\ln(\Lambda_2^3) = 0}$ (note that $\Lambda_1 = \Lambda_2$ since $m_1 = m_2$) for $T = 0.62$~{$\epsilon k_B^{-1}$}, and in doing so fix the energy scale by assigning a value to Planck's constant $h$ in Lennard-Jones units. This amounts to a constant shift in $\mu_i$ and does not affect $(\partial \mu_i / \partial x_1)_{P,T}$.

The excess chemical potential $\mu_i^{\mathrm{ex}}$ was calculated from MD simulations at constant $T$ and $P$ using a free energy perturbation (FEP) method. A single particle was inserted into the simulation cell via $(n-1)$ small ``steps'' along a reversible alchemical thermodynamic path, analogous to slowly ``growing'' the particle. The Gibbs free energy change for this particle insertion, $\Delta_N^{N+1} G^\mathrm{ex}$, is given by  
\begin{equation}
    \mu_i^\mathrm{ex} \approx \Delta_N^{N+1} G^\mathrm{ex} = 
    - k_B T \sum_{i=0}^{n-1}  \ln{\frac{ \langle V \exp{(-\Delta_{\zeta_i}^{\zeta_{i+1}} \mathcal{V} / k_B T)} \rangle_{\zeta_i} } {\langle V \rangle_{\zeta_i}}}	
\end{equation}
where $V$ is the volume of the simulation cell, $\Delta_{\zeta_i}^{\zeta_{i+1}} \mathcal{V}(\zeta, \bm{r}) = \mathcal{V}_{\zeta_{i+1}}(\zeta, \bm{r}) - \mathcal{V}_{\zeta_i}(\zeta, \bm{r})$ and $\zeta$ is a coupling parameter that connects the reference ($N$) and perturbed ($N+1$) systems according to $\mathcal{V}(\zeta, \bm{r}) = \zeta \mathcal{V}_{N+1}(\bm{r}) + (1- \zeta)\mathcal{V}_{N}(\bm{r})$, with $\zeta$ taking values from 0 to 1. In order to avoid singularities when $\zeta \rightarrow 0$, the FEP simulations were performed using a soft-core version\cite{Beutler1994_LJsoft} of the Lennard-Jones potential (LJSC) 
\begin{equation}\label{eq:LJsoft}
    \mathcal{V}_{ij}^\mathrm{LJSC}(r; \zeta, n, \alpha) = \zeta^n 4\epsilon_{ij} 
    \left\{ 
    \frac{1}{\left[ \alpha(1-\zeta)^2 + (r/\sigma_{ij})^6 \right]^2 }
    - \frac{1}{ \alpha(1-\zeta)^2 + (r/\sigma_{ij})^6}
    \right\}
\end{equation}
truncated and shifted at a cutoff radius of $r_c = 2.5 \sigma$ such that $\mathcal{V}_{ij}^\mathrm{LJSCTS}(r) = (\mathcal{V}_{ij}^\mathrm{LJSC}(r) - \mathcal{V}_{ij}^\mathrm{LJSC}(r_c)) \theta(r_c-r) $
with $\theta$ being the Heaviside step function. 
Values of $n=1$ and scaling constant $\alpha = 0.5$ were used. For $\zeta = 1$, eq.~\ref{eq:LJsoft} reduces to the standard Lennard-Jones potential, and in the limit $\zeta \rightarrow 0$ no work is required to change from $\alpha=0$ to $0 < \alpha < \infty$ (i.e. the two potentials have equivalent initial states). Therefore, the soft-core version gives the same free energy difference as the standard LJ (or LJTS) potential.\cite{Beutler1994_LJsoft}

Each FEP simulation used a cubic simulation cell containing 5000 particles, and additionally the single particle being inserted. The systems were equilibrated for $20 \tau$ in the \textit{NVT} ensemble, then for $2 \times 10^3 \tau$ in the \textit{NPT} ensemble. The particle insertion was then performed in steps of $0.01 \zeta$ over a $10^4 \tau$ production run. A timestep of $0.002\tau$ was used. 
Temperature (pressure) was controlled using the Nos\'{e}-Hoover chain thermostat (barostat), with 3 chains, and a time constant of $1\tau$ ($5\tau$).
Sampling consisted of 1700-2000 replicas for $\mu_1^\mathrm{ex}$, and 500-600 replicas for $\mu_2^\mathrm{ex}$.

\begin{figure}[t]
    \centering
    \begin{overpic}[width=1.0\textwidth,grid=False]{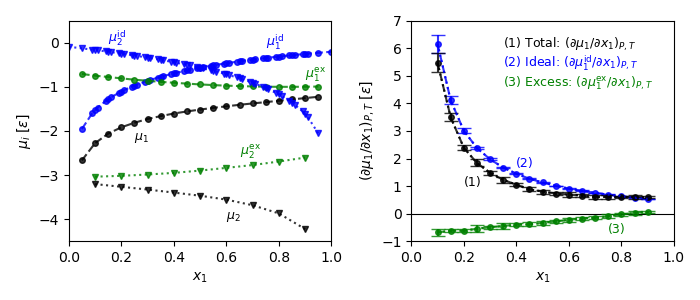}
        \put(0,40.5) {{\textsf{(a)}}}
        \put(50,40.5) {{\textsf{(b)}}}
    \end{overpic}
    \caption{The chemical potential and its derivative as a function of mole fraction $x_i$. (a) Total $\mu_i$, ideal $\mu_i^\mathrm{id}$ and excess $\mu_i^\mathrm{ex}$ chemical potentials $\mu_i$ of species $i=1,2$. (b) $(\partial \mu_1 / \partial x_1)_{P,T}$ and its ideal and excess contributions. In (a), statistical uncertainties are smaller than the size of the symbols.
    }
    \label{fig:chempot}
\end{figure}

The chemical potentials and the derivative $(\partial \mu_1 / \partial x_1)_{P,T} = (\partial \mu_1^\mathrm{id} / \partial x_1)_{P,T} + (\partial \mu_1^\mathrm{ex} / \partial x_1)_{P,T}$ are shown in {fig.~\ref{fig:chempot}}. $(\partial \mu_1^\mathrm{id} / \partial x_1)_{P,T}$ and $(\partial \mu_1^\mathrm{ex} / \partial x_1)_{P,T}$ were evaluated at $x_1$ by fitting a straight line through the three points within $x_1 \pm 0.01$ and $x_1 \pm 0.05$, respectively.

\subsection{Self-diffusion coefficients and shear viscosity}\label{subsec:diff_visc}

\begin{figure*}[!t]
    \centering
    \begin{overpic}[width=0.9\textwidth,grid=False]{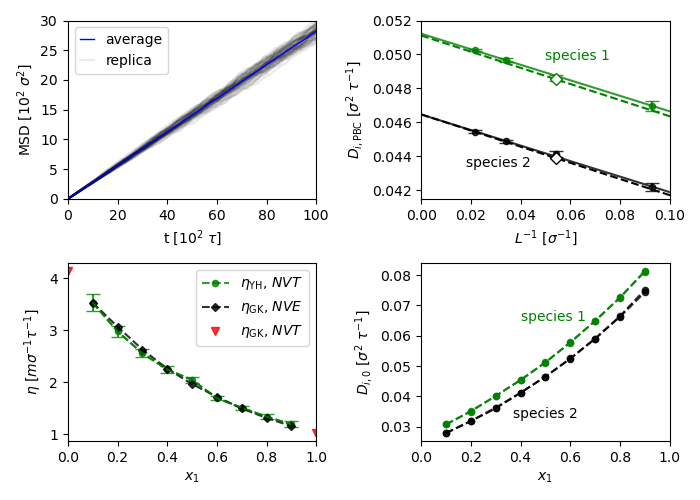}
        \put(1, 70) {{\textsf{(a)}}}
        \put(48, 70) {{\textsf{(b)}}}
        \put(1, 34) {{\textsf{(c)}}}
        \put(48, 34) {{\textsf{(d)}}}
    \end{overpic}
    \caption{Self-diffusion coefficients and shear viscosities of the Lennard-Jones mixtures. 
    (a) Mean-square displacement (MSD) \textit{vs.} time $t$ for the $x_1 = 0.5$ and $N=1000$ system.
    (b) Finite-size analysis for the $x_1 = 0.5$ mixture: $D_{i,\mathrm{PBC}}$ is the finite-size diffusion coefficient of species $i$ and $L$ is the length of the cubic simulation cell. Symbols: circles ($\circ$) and diamonds ($\diamond$) show the data from \textit{NVT} and \textit{NVE} MD simulations, respectively. Solid lines show a linear fit to the \textit{NVT} data; dashed lines show a straight line through the single \textit{NVE} data point with the gradient determined from $\eta_\mathrm{GK}$.     
    (c) The viscosities $\eta$ as a function of mole fraction $x_1$, as determined from the different methods (see main text).
    (d) ``Infinite-size'' diffusion coefficients $D_{i,0}$ of species $i$ as a function of $x_1$. $D_{i,0}$ from the two methods, \textit{NVT} simulations with direct extrapolation ($\eta_\mathrm{YH}$) and \textit{NVE} simulations with $\eta_\mathrm{GK}$, cannot be distinguished on the scale of the plot. 
    Where error bars are not explicitly shown, the statistical uncertainty is smaller than the size of the symbol.
    }
    \label{fig:diffFS}
\end{figure*}

The self-diffusion coefficients $D_i$ of species $i=1,2$ were calculated from the average mean square displacement (MSD) using the Einstein relation
\begin{equation}
    D_i = \frac{1}{2d} \lim_{t \to \infty} \frac{\langle | \bm{r}_i(t+t_0)-\bm{r}_i(t_0)|^2 \rangle}{t}
\end{equation} 
where $\bm{r}_i$ is the position vector of a particle of species $i$, $t$ is the elapsed time from arbitrary starting time $t_0$, and $d=3$ is the number of spatial dimensions. $D_i$ was therefore calculated by fitting to the equation $\langle | \bm{r}_i(t)-\bm{r}_i(0)|^2 \rangle = 6 t D_i$, excluding the first {10~{$\tau$}} of data to ensure only the diffusive regime was sampled (as opposed to the ballistic regime). 
In order to account for finite-size effects, the ``infinite-size'' diffusion coefficient $D_{i,0}$ was calculated by extrapolation to $L^{-1}=0$, where $L$ is the length of the cubic simulation cell. This finite-size analysis is shown in fig.~\ref{fig:diffFS}, and follows from the equation derived by Yeh and Hummer\cite{Yeh2004} using a simple hydrodynamic model of a particle surrounded by a solvent of viscosity $\eta_\mathrm{YH}$ in a periodically replicated simulation cell,
\begin{equation}\label{eq:YH_FS}
  D_{i,\mathrm{PBC}} = -\frac{\xi k_\mathrm{B} T}{6 \pi \eta_\mathrm{YH}} L^{-1} + D_{i,0}
\end{equation} 
where $D_{i,\mathrm{PBC}}$ are the finite-size diffusion coefficients calculated from our MD simulations, and $\xi$ is a dimensionless constant equal to 2.837297 for a cubic simulation cell with 3D periodic boundary conditions.\cite{Yeh2004} The same expression was obtained earlier\cite{Dunweg1993} by D\"{u}nweg and Kremer using a closely related derivation. For each state point, the $i=1$ and $i=2$ data sets were simultaneously fit to eq.~\ref{eq:YH_FS}, giving a single viscosity $\eta_\mathrm{YH}$ for the mixture.

\begin{figure*}[!t]
    \centering
    \begin{overpic}[width=0.8\textwidth,grid=False]{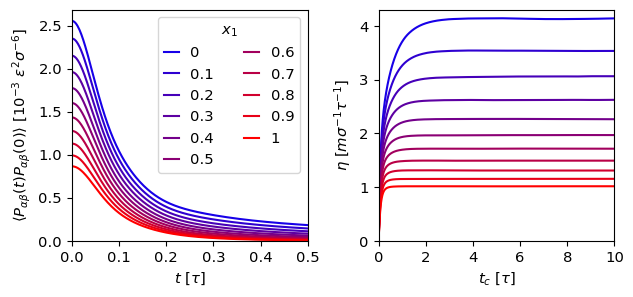}
        \put(0, 46) {{\textsf{(a)}}}
        \put(51, 46) {{\textsf{(b)}}}
    \end{overpic}
    \caption{(a) Autocorrelation functions $\langle P_{\alpha \beta}(t) P_{\alpha \beta}(0) \rangle$ as a function of time lag $t$. 
    (b) Convergence of shear viscosity $\eta(t_c)$ with the correlation time $t_c$ used as the upper limit of the integral in {eq.~\ref{eq:GKvisc}}.
      }
    \label{fig:viscCF}
\end{figure*}

The shear viscosity $\eta$ was additionally calculated using the Green-Kubo integral formula
\begin{equation}\label{eq:GKvisc}
    \eta_{GK} = \frac{V}{k_\mathrm{B} T} \lim_{t^\prime \to \infty} \int_0^{t^\prime} \langle {P_{\alpha \beta}(t) P_{\alpha \beta}(0)} \rangle dt
\end{equation}
where $\alpha \neq \beta$ such that $P_{\alpha \beta}$ are the off-diagonal elements of the pressure tensor. Results were averaged over $(\alpha, \beta) = (x,y)$, $(x,z)$ and $(y,z)$. A correlation time of {$t_c = 10 \tau$} was used for the upper limit of the integral, which is sufficient to obtain well-converged integrals (see {fig.~\ref{fig:viscCF}}). 

In order to accurately target a specific thermodynamic states when calculating $D_{i,\mathrm{PBC}}$, the \textit{NVT} ensemble was sampled using a temperature-control algorithm that alters dynamics compared to the Newtonian dynamics of the \textit{NVE} ensemble. It is therefore necessary to check whether the employed global Nosé–Hoover (three chains) thermostat affects the computed self-diffusion coefficients. We show in {fig.~\ref{fig:diffFS}(b)\&(c)} that it does not: all diffusion coefficients and viscosities calculated from \textit{NVT} and \textit{NVE} simulations, and using the different methods (eqs.~\ref{eq:YH_FS}\&\ref{eq:GKvisc}) agree to within their associated uncertainties.  
This is consistent with previous work that shows that “global” velocity
scaling thermostats, including the Nosé–Hoover-chain thermostat, do not
significantly alter diffusion coefficients or viscosity.~\cite{Basconi2013}

\subsection{Kirkwood-Buff integrals from grand canonical Monte Carlo}

KBIs were calculated from grand canonical Monte Carlo simulations. 
For each simulation, the temperature of the ideal gas reservoir was set to {$T = 0.62$}~{$\epsilon k_B^{-1}$}, and the input chemical potentials $\mu_1$ and $\mu_2$ were determined from the free-energy perturbation simulations described in {sec.~\ref{subsec:FEP}}. In all cases, a cubic simulation cell with length $L=20 \sigma$ was used.  
Each MC step, 100 trial displacement and 100 trial exchanges (insertions or deletions with equal probability) were attempted. For the displacement moves, a maximum translation distance of $1.0 \sigma$ was allowed. Trial moves were accepted/rejected using the standard Metropolis criterion. Each replica was first equilibrated in an \textit{NVT} MD simulation at the target $x_1$ and $\rho_N$ for {200~$\tau$}, followed by $10^5$ MC steps, and finally a production run of $10^7$ MC steps. Sampling consisted of 400-500 statistically independent replicas for each state point.      

The Kirkwood-Buff integrals $G_{ij}$ were calculated from the particle number fluctuations according to 
\begin{equation}
    G_{ij} = V\frac{\langle N_i N_j \rangle - {\langle N_i \rangle}{\langle N_j \rangle}}{{\langle N_i \rangle}{\langle N_j \rangle}} 
    - V\frac{\delta_{ij}}{\langle N_j \rangle}
\end{equation}
where $N_i$ is the number of particles of species $i$ and $V$ is the volume of the simulation cell. The average pressures are within $\pm 0.008 \epsilon \sigma^{-3}$ of {$P = 0.46$~{$\epsilon \sigma^{-3}$}}, corresponding to differences in number density on the order of $\delta \rho_N \sim 10^{-4} \sigma^{-3}$ relative to those calculated from \textit{NPT} MD simulations (see fig.~\ref{fig:EOSandPD}(a)). Differences in $G_{ij}$ and subsequently $\Gamma$ due to the slightly different thermodynamic states being sampled are expected to be insignificant compared to their associated statistical uncertainties.

\subsection{Kirkwood-Buff integrals from molecular dynamics (\textit{NVT}) simulations}\label{sec:KBI_RDF}

For an infinitely large and open three-dimensional system, the KBI for mixture components $i$ and $j$ is defined as
\begin{equation}\label{eq:KBIinf}
    G_{ij} = 4 \pi {\int_{0}^{\infty} [g_{ij}^{\mu V T}(r) - 1] r^2 dr}
\end{equation}
where $r$ is the radial distance between particles $i$ and $j$.
The infinite-size KBIs were calculated from a finite-size analysis of finite-volume KBIs 
\begin{equation}
    G_{ij}^V = \int_V [g_{ij}^{\mu V T}(r) -1 ]w(r) dr
\end{equation}
defined for finite and open subvolumes embedded in a reservoir.\cite{Kruger2013_FiniteVolumeKBI} 
For spherical symmetry $w(r) = 4 \pi r^2 (1 - 3x/2 + x^3/2)$ valid for $x < 1$, where $x=r/(2R)$ and $R$ is the radius of the subvolume. $g_{ij}^{\mu V T}$ is the pair correlation function \textit{in the thermodynamic limit}.
It has been shown that $G_{ij}^V$ scales linearly with $R^{-1}$ and that the infinite-size $G_{ij}$ can be obtained by extrapolating the linear regime to $R^{-1} = 0$.\cite{Kruger2013_FiniteVolumeKBI,Dawass2018_FiniteSizeEffectsKBI,Ganguly2013_KBIcorrection}

\begin{figure*}[t]
    \centering
    \begin{overpic}[width=1.0\textwidth,grid=False]{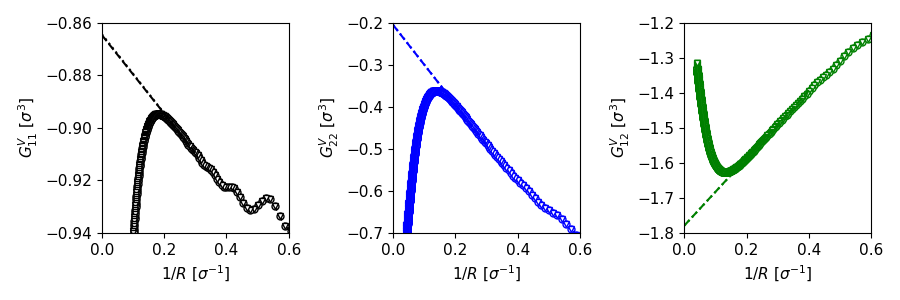}
        \put(28, 27.5) {{\textsf{(a)}}}
        \put(60, 27.5) {{\textsf{(b)}}}
        \put(92.5, 9) {{\textsf{(c)}}}
    \end{overpic}
    \caption{Extrapolation of the finite-volume Kirkwood-Buff integrals $G_{ij}^V$ to $R^{-1} =0$, where $R$ is the radius of the subvolume: (a) $G_{11}^V$, (b) $G_{22}^V$ and (c) $G_{12}^V$.  The data shown corresponds to mole fraction $x_1 = 0.5$ and a system size of $N=5 \times 10^6$ particles. Symbols: circles ($\cdot \cdot \circ \cdot \cdot$) and downward triangles ($\cdot \cdot \triangledown \cdot \cdot$) denote the use of pair correlation function $g_{ij}$ (uncorrected) and the corrected $g_{ij}^{GV}$, respectively, although these cannot be easily distinguished on the scale of the plots.  }
    \label{fig:KBIextrap}
\end{figure*}

We calculate the RDFs from simulations of $N=10^5$ to $N=5 \times 10^6$ particles in the \textit{NVT} ensemble. Because fluctuations transform between ensembles, $g_{ij}^{\mu V T}$ cannot be formally replaced with $g_{ij}^{NVT}$. However, for a sufficiently large system, the correlation lengths of particle density fluctuations are small compared to the linear dimension of the simulation cell, and local correlations are expected to be well reproduced. In addition to a large system size, we apply the tail correction\cite{Ganguly2013_KBIcorrection} of {Ganguly and van der Vegt} to ensure the correct asymptotic limit ($\lim_{r \to\infty} g_{ij} = 1$) of the RDF. 
\begin{equation}
    g_{ij}^{GV}(r) =  g_{ij}(r)  
                      \left( \frac{N_j h(r)}{N_j h(r) - n_{ij}^{ex}(r) - \delta_{ij}} \right)
\end{equation}
\begin{equation}
    h(r) = 1 - \frac{4 \pi r^3}{3 V}
\end{equation}
where $N_j$ is the total number of particles $j$ in the simulation cell of volume $V$. However, we note that in our simulations, this correction does not have an appreciable effect on the linear regime and subsequently the extrapolated $G_{ij}$ (see fig.~\ref{fig:KBIextrap}). Other corrections\cite{Salacuse1996_FiniteSizeEffectsRDF,CortesHuerto2016_KBIcorrection} to the RDFs for calculating KBIs have been proposed; the correction of {Ganguly and van der Vegt} was found to be the most accurate for a WCA system, relative to a larger reference system ($L = 80 \sigma$).\cite{Dawass2018_FiniteSizeEffectsKBI}

\begin{figure*}[t]
    \centering
    \begin{overpic}[width=1.0\textwidth,grid=False]{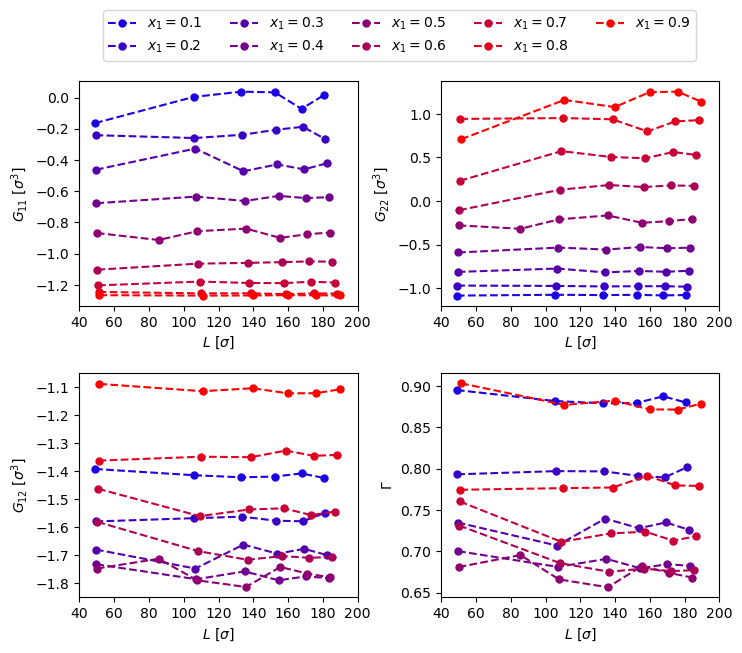}
        \put(0, 77) {{\textsf{(a)}}}
        \put(50, 77) {{\textsf{(b)}}}
        \put(0, 38) {{\textsf{(c)}}}
        \put(50, 38) {{\textsf{(d)}}}
    \end{overpic}
    \caption{Finite-size analysis of the infinite-size Kirkwood-Buff integrals $G_{ij}$ as a function of simulation cell length $L$, for different mole fractions $x_1$. (a) $G_{11}$, (b) $G_{22}$, (c) $G_{12}$, and (d) thermodynamic factor $\Gamma$.
    }
    \label{fig:KBI_FS}
\end{figure*}

We show in {fig.~\ref{fig:KBI_FS}} a finite-size analysis of the infinite-size KBIs. The final values for $G_{ij}$ in the main text were taken from the largest system size, and the associated uncertainties were estimated from the convergence with system size.

\subsection{Melting points of the pure components}\label{subsec:melting_points}

The melting temperatures $T_m$ of the pure components $i=1,2$ were determined using the direct coexistence method in the $NP_zT$ ensemble. This involves preparing a solid-liquid interface at a given ($T,P$) state point and observing whether the system completely melts, freezes, or remains in coexistence. 
In MD simulations, cooling a homogeneous liquid below its melting point usually results in a metastable supercooled liquid; freezing is not observed unless prohibitively long simulations are performed. Likewise, heating a solid slightly above its melting point results in a superheated solid. The presence of the interface lowers the kinetic barrier for melting/freezing. 

First, the densities of the solid and liquid phases were determined from \textit{NPT} simulations. Initial configurations were prepared by joining two half-boxes, one of the FCC-solid the other of the liquid, each containing {$N=5324$} particles and equilibrated in the \textit{NVT} ensemble. The combined {$N=10648$} system was then equilibrated for {$20 \tau$}, ensuring that the solid phase did not melt by restraining each solid particle to its equilibrium position $\bm{r}_0$ with a harmonic potential, $\mathcal{V}(\bm{r})= k(\bm{r} - \bm{r_0})^2/2 $ of force constant {$k = 10$~$\epsilon \sigma^{-2}$}. A {$4 \times 10^4 \tau$} production run in the $NP_zT$ ensemble was then performed, applying a barostat only to the direction parallel to the surface normal (the $z$-direction). A Nos\'{e}-Hoover chain barostat was used, with 3 chains, and a time constant of {4~$\tau$}. Likewise, a Nos\'{e}-Hoover chain thermostat was used, also with 3 chains, and a time constant of {1~$\tau$}. Sampling consisted of an initial 10 statistically independent replicas, followed by an additional 40 replicas if the initial set did not all either melt or freeze. In all cases, a homogeneous phase was observed by the end of the production run. 

The simulations were performed in the $NP_zT$ ensemble, and as such the simulation cell dimensions ($L_x$ and $L_y$) were chosen to be consistent with the density of the solid at the ($T,P$) state point. Otherwise, the solid phase would possess internal stress,\cite{Espinosa2013_meltingPHS,Frenkel2013_darkside} corresponding to a higher free energy, and resulting in the overestimation of the coexistence pressure and/or underestimation of the melting temperature. 

\begin{figure*}[!t]
    \centering
    \includegraphics[width=0.74\textwidth]{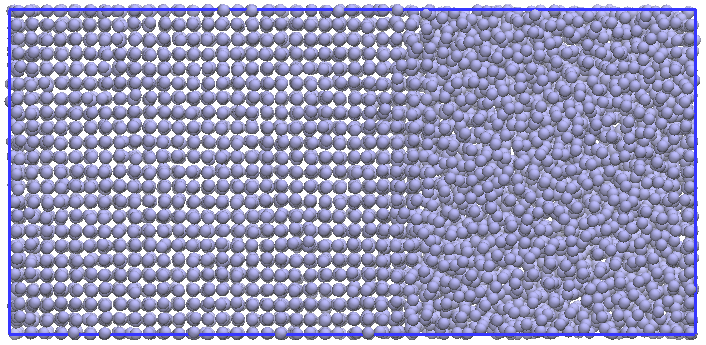}
    \par \vspace{0.1cm}
    \begin{overpic}[width=0.95\textwidth,grid=False]{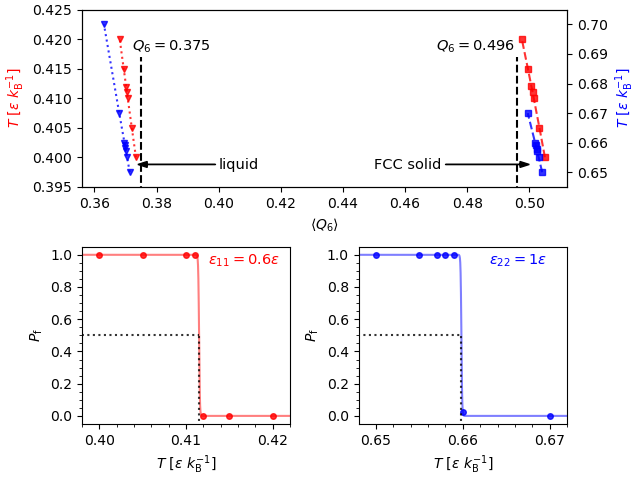}
        \put(-1, 113) {{\textsf{(a)}}}
        \put(-1, 73) {{\textsf{(b)}}}
        \put(-1, 36) {{\textsf{(c)}}}
    \end{overpic}
    \caption{Melting points of the pure components determined using the direct coexistence method. (a) Snapshot of the simulation cell in the process of melting/freezing. (b) Lennard-Jones liquids and FCC solids at temperatures $T$ characterized by their average $Q_6$ order parameter. Species 1 (species 2) is shown in red (blue) and on the left (right) axis. (c) Probability of freezing $P_f$ as a function of temperature $T$ for species 1 (left) and 2 (right). Solid lines show the fitted sigmoid-like functions (eq.~\ref{eq:Prob_freeze}). 
    }
    \label{fig:melting}
\end{figure*}

Each replica was determined to have either frozen or melted using the $Q_6$ ($l=6$) bond-orientational order parameter. The Steinhardt order parameters $Q_l$ were introduced to characterize local orientational order in atomic structures,\cite{Steinhardt1983_BondOrientationalOP} and are given by
\begin{equation}
    Q_l = \sqrt{ \frac{4 \pi}{2l +1} \sum_{m=-l}^{+l} {\bar{Y}_{lm} \bar{Y}^*_{lm}} }
\end{equation}
\begin{equation}
    \bar{Y}_{lm} = \frac{1}{N_n} \sum_{j=1}^{N_n} {Y_{lm}(\theta (\bm{r}_{ij}), \phi (\bm{r}_{ij}) )}
\end{equation}
for each particle $i$, where ${Y_{lm}(\theta, \phi)}$ are the spherical harmonics, $\theta$ and $\phi$ are the polar angles of "bond" vector $\bm{r}_{ij}$ between $i$ and neighbour $j$, and $N_n$ is the number of nearest neighbours to particle $i$. $Q_l$ is therefore a rotationally invariant non-negative amplitude. 
$Q_l$ adopt well-defined values for high-symmetry structures; for a perfect FCC crystal ($Fm\bar{3}m$) and $N_n = 12$, $Q_6 = 0.575$.\cite{Mickel2013_BondOrientationalOP}     
We show in fig.~\ref{fig:melting} that the $Q_6$ order parameter can be used to distinguish between the FCC solid and liquid phases. The replica was determined to have frozen if $\langle Q_6 \rangle > 0.496 $, or melted if $\langle Q_6 \rangle < 0.375$. The average $\langle Q_6 \rangle$ was taken over the last {10~$\tau$} of the trajectory.

In the thermodynamic limit, a solid melts at $T > T_m$, and a liquid freezes for $T < T_m$. However, a finite-size system may stochastically melt or freeze with probabilities $P_m$ and $P_f = 1 - P_m$ respectively. 
For each component, $T_m$ was determined by fitting $P_f(T)$ to the sigmoid-like function (fig.~\ref{fig:melting}):
\begin{equation}\label{eq:Prob_freeze}
    P_f(T) = \frac{1}{2} - \frac{1}{2} \tanh{ \left[ (T - T_m)/d \right]}
\end{equation}
where $d$ controls the sharpness of the probability profile, with $\delta_{10-90} = 2.178 d$ being the width of the interval where $P_f$ goes from 0.1 to 0.9. The melting (coexistence) temperature is defined by $P_m(T_m) = P_f(T_m) = 0.5$. The melting temperatures are shown in table~\ref{table:melting} alongside other coexistence properties (coexistence densities and enthalpies were determined by interpolating data from the \textit{NPT} simulations).

\begin{table*}[h!]
\centering
\caption{Solid-liquid coexistence properties of the pure LJ components at {$P = 0.46$~{$\epsilon \sigma^{-3}$}}: the melting temperature $T_m$; coexistence number densities of the FCC solid and liquid phases, $\rho_{N,s}$ and $\rho_{N,l}$, respectively; and the enthalpy of fusion $\Delta H_\mathrm{fus}$}.
\label{table:melting}
\begin{threeparttable}
\begin{tabular}{c c c c c}
\toprule
\hline 
 Species & $T_m$ [{$\epsilon k_\mathrm{B}^{-1}$}] 
 & $\rho_{N,s}$ [$\sigma^{-3}$] & $\rho_{N,l}$ [$\sigma^{-3}$] 
 & $\Delta H_\mathrm{fus}$ [$\epsilon$] \\

\hline
 1 & 0.4115(5) & 0.9550(2) & 0.8500(3) & 0.523(3) \\
\hline
 2 & 0.6598(3) & 0.9509(1) & 0.8419(2) & 0.927(3) \\

\hline
\end{tabular}

\end{threeparttable}
\end{table*}

\newpage
\printbibliography
